\documentclass[aps,prd,reprint,nofootinbib,amsmath,amssymb,longbibliography,floatfix]{revtex4-2}
\usepackage[T1]{fontenc}
\usepackage{lmodern}
\usepackage{graphicx,bm,booktabs}
\usepackage{multirow}
\usepackage[colorlinks=true,linkcolor=blue,citecolor=blue,urlcolor=blue]{hyperref}
\hypersetup{pdftitle={Construction of f4 family},pdfsubject={Ground-state-calibrated pair-creation analysis at gamma=6.90}}
\graphicspath{{figures/}}

\newcommand{\dd}{\mathrm{d}}

\usepackage{xcolor}
\definecolor{F4RevisionRed}{RGB}{185,20,35}

\begin{document}
\title{Construction of the $f_4$ meson family}
\author{
Zhe-Qi Huang$^{1,2,3}$}
\author{
Lei Huang$^{4}$}
\author{
Yu-Qiang Li$^{2}$}
\author{
Xin-Lei Ge$^{1}$}\email{gexl\_bhu@163.com}

\author{
Hao Chen$^{2,5}$}\email{chenhao_qhnu@outlook.com
}
\author{Cheng-Qun Pang$^{3,6}$}\email{xuehua45@163.com}
\affiliation{
$^1$ 
College of Physical Science and Technology,
Bohai University, Jinzhou 121013, China
\\
$^2$ 
College of Physics and Electronic Information Engineering,
Qinghai Normal University, Xining 810000, China
\\
$^3$ School of Physics and Optoelectronic Engineering,
Ludong University, Yantai 264000, China\\
$^4$ Heze Medical College, Heze 274000, China\\
$^5$ 
Academy of Plateau Science and Sustainability,
Xining 810016, China
\\
$^6$ 
Lanzhou Center for Theoretical Physics,
Key Laboratory of Quantum Theory and Applications of MoE,
and Key Laboratory of Theoretical Physics of Gansu Province,
Lanzhou University, Lanzhou, Gansu 730000, China
}
\date{October 13, 2026}
\begin{abstract}

Motivated by the experimental observations of $f_4(2050)$ and $f_4(2300)$ and the establishment of $a_4$ family, we investigate the $f_4$ family by combining the mass spectrum under the modified Godfrey--Isgur model with two-body strong-decay calculations in the quark-pair-creation model.
The common pair-creation strength is fixed by the measured total width of $f_4(2050)$.
The mass spectrum analysis and decay information both favor the nonstrange $2\,{}^3F_4$ assignment for $f_4(2300)$.
The largely strange $2F$ partner is predicted at $2453$ MeV with a total width of $216$ MeV, decaying mainly into $K^*\bar K^*(1410)$ and $K^*\bar K^*$.
Predictions are also given for the $3F$, $4F$, $1H$, and $2H$ flavor pairs.
At the adopted mixing angles, the $H$-wave states are more than twice as broad as nearby $F$-wave states of the same flavor.
The results of this work may serve as useful guidance for future experimental studies.
\end{abstract}

\maketitle

\section{Introduction}\label{sec1}
Light mesons provide a useful probe of the interplay among orbital excitations, flavor structure, and strong-decay dynamics in the nonperturbative regime of quantum chromodynamics (QCD)~\cite{Godfrey:1998pd}.
Experimental studies have identified $f_4(2050)$ and the higher-mass candidate $f_4(2300)$, but their spectroscopic relationship and the identification of the strange partner remain unresolved~\cite{Anisovich:2000ut,WA102:2000bii,Ebert:2009ub,Chen:2021kfw}.
Taking the experimentally established $f_4(2050)$ as a reference and drawing on recent theoretical work on the isovector $a_4$ family~\cite{Wang:2025hve}, we investigate the mass spectrum and strong decays of the $f_4$ family which can help complete the light-meson spectrum.

The $f_4$ family is characterized by the quantum numbers $I^G(J^{PC})=0^+(4^{++})$.
In the conventional $q\bar q$ picture, these quantum numbers correspond to a quark--antiquark pair with total spin $S=1$ and total angular momentum $J=4$.
The parity and angular-momentum coupling constraints allow the relative orbital angular momentum to be either $L=3$ or $5$, corresponding to the $n\,{}^3F_4$ and $n\,{}^3H_4$ configurations, respectively.
Unlike the isovector $a_4$ family, the isoscalar $f_4$ sector contains both the nonstrange component $|n\bar n\rangle=(|u\bar u\rangle+|d\bar d\rangle)/\sqrt{2}$ and the strange component $|s\bar s\rangle$.
Constructing this family therefore requires distinguishing radial from orbital excitations and examining the flavor composition and mixing of the candidate states~\cite{Godfrey:1985xj,Anisovich:2000kxa,Ebert:2009ub,Feng:2012fma}.

Early experimental studies of $f_4(2050)$ date back to 1975.
The CERN--MUNICH Collaboration observed a spin-four resonance with a mass of approximately $2050\,\mathrm{MeV}$ in a partial-wave analysis of $\pi^-p\to K^+K^-n$ at a beam momentum of $18.4\,\mathrm{GeV}/c$~\cite{CERN-MUNICH:1975yng}.
In the same year, the Serpukhov--CERN Collaboration observed a structure with a mass of $2020\pm30\,\mathrm{MeV}$ in $\pi^-p\to\pi^0\pi^0n$ at $40\,\mathrm{GeV}/c$.
The decay angular distributions and higher-order angular moments supported a spin-four assignment~\cite{Serpukhov-CERN:1975qnq}.
The resonance observed in these experiments, then known as the $h$ meson, is now listed as $f_4(2050)$ by the Particle Data Group (PDG)~\cite{ParticleDataGroup:2026mpi}.
In 1982, the Notre Dame--Argonne group used $\pi^+p\to\Delta^{++}\pi^0\pi^0$ data at a beam momentum of $8.0\,\mathrm{GeV}/c$ to extract the $\pi^+\pi^-\to\pi^0\pi^0$ scattering amplitudes by extrapolating the angular moments to the pion pole.
A Breit--Wigner fit to the $G$-wave amplitude yielded the mass, width, and $\pi\pi$ branching fraction of this state~\cite{Cason:1982xx}.

In 2000, a combined analysis used Crystal Barrel data together with charged-dipion data to study $\bar pp\to\eta\pi^0\pi^0$, $\pi^0\pi^0$, $\eta\eta$, $\eta\eta'$, and $\pi^+\pi^-$.
The analysis identified an isoscalar $J^{PC}=4^{++}$ resonance with a mass of $2283\pm17\,\mathrm{MeV}$ and a width of $310\pm25\,\mathrm{MeV}$~\cite{Anisovich:2000ut}.
In the same year, the WA102 Collaboration performed a spin analysis of the $\omega\omega$ system in the central-production reaction $pp\to p_f(\omega\omega)p_s$ at a beam momentum of $450\,\mathrm{GeV}/c$, in which a Breit--Wigner fit to the $J^{PC}=4^{++}$ wave yielded a mass of $2332\pm15\,\mathrm{MeV}$ and a width of $260\pm57\,\mathrm{MeV}$~\cite{WA102:2000bii}.
In a study of the $\eta\pi^+\pi^-$ system in the reaction $\pi^-p\to\eta\pi^+\pi^-n$ by the VES Collaboration, the intensities and relative phases of the $4^{++}$ partial waves in the $a_2(1320)\pi$ and $f_2(1270)\eta$ channels provided evidence for a higher-mass resonance, with the signal appearing mainly in $f_2(1270)\eta$~\cite{VES:2000xjq}.
These results are included in the PDG listing for $f_4(2300)$~\cite{ParticleDataGroup:2026mpi}.
Selected experimental masses and widths are summarized in Table~\ref{tab:1}.

The $f_4(2050)$ is commonly regarded as a largely nonstrange $1\,{}^3F_4$ state in theoretical studies.
The Godfrey--Isgur (GI) relativized quark model predicts the nonstrange $1\,{}^3F_4$ level at approximately $2010\,\mathrm{MeV}$~\cite{Godfrey:1985xj}, while the relativistic quasipotential model of Ebert \textit{et al.} gives a corresponding mass of $2018\,\mathrm{MeV}$~\cite{Ebert:2009ub}.
Feng and Wei studied Regge relations and flavor mixing in the $1\,{}^3F_4$ nonet; in their mixing scheme, $f_4(2050)$ is almost entirely nonstrange~\cite{Feng:2012fma}.

By contrast, the assignment of $f_4(2300)$ remains unsettled.
Ebert \textit{et al.} predicted nearly degenerate masses of $2284$ and $2286\,\mathrm{MeV}$ for the nonstrange $2\,{}^3F_4$ and strange $1\,{}^3F_4$ states, respectively~\cite{Ebert:2009ub}.
Our previous work supported the interpretation of $f_4(2300)$ as a nonstrange $2\,{}^3F_4$ state through ${}^3P_0$ decay calculations with harmonic-oscillator wave functions~\cite{Pang:2015eha}.
In a Regge-trajectory analysis, however, Chen and Masjuan tentatively placed $f_4(2300)$ on the $s\bar s$ trajectory as a $1\,{}^3F_4(s\bar s)$ candidate~\cite{Chen:2021kfw,Masjuan:2012gc}.
In two studies of the $1\,{}^3F_4$ nonet, Feng and Wei obtained masses of $2167\pm27\,\mathrm{MeV}$ and $2129\pm20\,\mathrm{MeV}$ for the strange partner~\cite{Feng:2012fma,Feng:2023tue}, both below the experimental mass of $f_4(2300)$~\cite{ParticleDataGroup:2026mpi}.
The latter study therefore disfavors assigning $f_4(2300)$ to the $1\,{}^3F_4(s\bar s)$ state~\cite{Feng:2023tue}.
These competing assignments motivate a joint analysis of the mass spectrum and strong-decay properties.

Lattice QCD provides a first-principles nonperturbative framework~\cite{Dudek:2011tt,Dudek:2013yja}, but determining the properties of highly excited resonances with multiple open decay channels remains challenging~\cite{Briceno:2017max}.
QCD sum rules~\cite{Shifman:1978bx}, quark potential models~\cite{Godfrey:1985xj}, and phenomenological decay models~\cite{Barnes:1996ff} remain important tools for investigating meson spectra and decay properties.

\begin{table}[t]
\caption{Experimental information on the $f_4$ states.
Masses and widths are in MeV.
The $f_4(2050)$ row gives the PDG averages~\cite{ParticleDataGroup:2026mpi}.
For $f_4(2300)$, the first three rows give individual determinations, and the last row gives the PDG estimates.}
\label{tab:1}
\centering
\begin{tabular}{@{}lcc@{}}
\toprule
State & Mass & Width \\
\midrule
$f_4(2050)$ & $2018\pm11$ & $237\pm18$~\cite{ParticleDataGroup:2026mpi} \\
\addlinespace[6pt]
\multirow{4}{*}{$f_4(2300)$}
 & $2283\pm17$ & $310\pm25$~\cite{Anisovich:2000ut} \\
 & $2332\pm15$ & $260\pm57$~\cite{WA102:2000bii} \\
 & $2330\pm20\pm40$ & $235\pm50\pm40$~\cite{VES:2000xjq} \\
 & $2320\pm60$ & $250\pm80$~\cite{ParticleDataGroup:2026mpi} \\
\bottomrule
\end{tabular}
\end{table}

Among quark potential models, the GI model has been widely used to describe meson spectra and to obtain the corresponding spatial wave functions~\cite{Godfrey:1985xj}.
For higher radial and orbital excitations, however, the treatment of the long-range confining interaction affects both the energy levels and the wave functions~\cite{Song:2015fha}.
To incorporate color-screening effects phenomenologically, the modified Godfrey--Isgur (MGI) model replaces the linear confining potential of the GI model with a screened form.
Song \textit{et al.} applied this treatment to charmed-strange mesons and combined the resulting spectrum with strong-decay calculations to examine candidate assignments~\cite{Song:2015nia}.
The combination of the MGI and quark-pair-creation (QPC) models has also been applied to the kaon family~\cite{Pang:2017dlw} as well as to the $J^{PC}=3^{++}$~\cite{Chen:2026mdo}, $J^{PC}=4^{++}$~\cite{Wang:2025hve} (isovector), and $5^{++}$~\cite{Pang:2018gcn} light-meson families.

For two-body strong decays, we adopt the QPC model, also known as the ${}^3P_0$ model.
In this framework, a quark--antiquark pair with the vacuum quantum numbers $J^{PC}=0^{++}$ is created, and its constituents recombine with those of the initial meson to form two final-state mesons.
The model is widely used to calculate the widths of Okubo--Zweig--Iizuka (OZI)-allowed strong decays~\cite{Micu:1968mk,LeYaouanc:1972vsx,Ackleh:1996yt,Barnes:1996ff,Barnes:2002mu}.

In this work, we combine the MGI mass spectrum and numerical spatial wave functions with QPC decay calculations.
Taking $f_4(2050)$ as the nonstrange $1\,{}^3F_4$ reference state, we use its measured total width to determine the common pair-creation strength and compare its relative decay rates with experiment.
We then examine the mainly nonstrange $2\,{}^3F_4$ and mainly strange $1\,{}^3F_4$ assignments of $f_4(2300)$.
We also predict the decay properties of the strange $2F$ partner and the $3F$, $4F$, $1H$, and $2H$ flavor pairs, and examine their dependence on nonstrange--strange flavor mixing.

The paper is organized as follows.
Section~\ref{sec2} introduces the MGI and QPC models and specifies the normalization and flavor conventions.
In Sec.~\ref{sec3}, we present our numerical results, discuss the assignments of $f_4(2050)$ and $f_4(2300)$, and give predictions for higher excitations and their strange partners.
Conclusions are given in Sec.~\ref{sec4}.

\section{Models employed in this work}\label{sec2}

\subsection{The modified Godfrey--Isgur model}

The MGI model retains the relativistic treatment of the original GI model and replaces the linear confining potential with a screened form~\cite{Godfrey:1985xj,Song:2015nia}.
In the center-of-mass frame of the meson, its effective Hamiltonian is written as~\cite{Godfrey:1985xj,Li:2023wgq,Song:2015nia}
\begin{equation}
\label{eq:MGI_Hamiltonian}
 \tilde{H}
 =
 \sqrt{\mathbf{p}^{2}+m_1^{2}}
 +
 \sqrt{\mathbf{p}^{2}+m_2^{2}}
 +
 \tilde{V}^{\mathrm{eff}}(\mathbf{p},\mathbf{r}),
\end{equation}
where parameters $m_1$ and $m_2$ denote the constituent masses of the quark and antiquark, $\mathbf{p}$ is the three-momentum of either constituent, and
\begin{equation}
 \tilde{V}^{\mathrm{eff}}(\mathbf{p},\mathbf{r}) =
 \tilde{V}^{\mathrm{conf}}
 + \tilde{V}^{\mathrm{hyp}}
 + \tilde{V}^{\mathrm{so}}
\end{equation}
denotes the effective quark--antiquark interaction.
The tildes indicate relativized interaction terms.
The relativization procedure involves smearing the coordinate-space potentials and introducing momentum-dependent correction factors.
The potential before relativization is written as
\begin{equation}
\label{eq:V_eff_nonrelativistic}
 V^{\mathrm{eff}}(r) = V^{\mathrm{conf}}(r) + V^{\mathrm{hyp}}(r) + V^{\mathrm{so}}(r).
\end{equation}
The spin-independent potential $V^{\mathrm{conf}}(r)$ is given by
\begin{equation}
 V^{\mathrm{conf}}(r) = S(r) + G(r),
\end{equation}
where
\begin{equation}
 S(r) = \frac{b}{\mu}\left(1-e^{-\mu r}\right) + c
\end{equation}
is the screened scalar confining potential, and
\begin{equation}
 G(r) = -\frac{4\alpha_s(r)}{3r}
\end{equation}
is the Coulomb-like potential.
Here, $\alpha_s(r)$ is the running strong coupling in coordinate space

\begin{equation}
\alpha_s(r)=\sum_k\alpha_k\left(\frac{2}{\sqrt{\pi}}\int_0^{\gamma_{k}r} e^{-x^2}dx\right),
\end{equation}
with $\alpha_k=(0.25, 0.15, 0.2)$ and $\gamma_k=(1/2{\,\text{GeV}},\sqrt{10}/2{\,\text{GeV}},\sqrt{1000}/2{\,\text{GeV}})$ for $k=1,2,3$~\cite{Godfrey:1985xj}.
The parameter $b$ denotes the string tension and $c$ is a constant.
Their values along with the other MGI parameters used in this work are given by Ref.~\cite{Wang:2024lba} and listed in Table~\ref{tab:parameters}.
The screening parameter $\mu$ sets the characteristic length scale $1/\mu$.
In the limit $\mu\to0$, $S(r)$ reduces to the linear confining potential $br+c$ of the original GI model.
At large distances, it approaches the constant $b/\mu+c$.

The hyperfine potential $V^{\mathrm{hyp}}(r)$ consists of contact and tensor terms,
\begin{equation}
V^{\mathrm{hyp}}(r)
=
V^{\mathrm{cont}}(r)
+
V^{\mathrm{tens}}(r).
\end{equation}
The contact interaction takes the form
\begin{equation}
\begin{split}
{V}^{\mathrm{cont}}(r)=\frac{2{\bf S}_1\cdot{\bf S}_2}{3m_1m_2}\nabla^2{G}(r),
\end{split}
\end{equation}
where $\mathbf{S}_1$ and $\mathbf{S}_2$ denote the spin operators of the quark and antiquark, respectively.

The tensor interaction is written as
\begin{equation}
\begin{aligned}
V^{\mathrm{tens}}(r)
={}&
-\frac{1}{3m_1m_2}
\left[
3\frac{
(\mathbf{S}_1\cdot\mathbf{r})
(\mathbf{S}_2\cdot\mathbf{r})
}{r^2}
-\mathbf{S}_1\cdot\mathbf{S}_2
\right]
\\
&\times
\left(
\frac{\mathrm{d}^2}{\mathrm{d}r^2}
-\frac{1}{r}\frac{\mathrm{d}}{\mathrm{d}r}
\right)G(r).
\end{aligned}
\label{Vtens}
\end{equation}

The spin--orbit interaction $V^{\mathrm{so}}(r)$ consists of vector and scalar contributions,
\begin{equation}
V^{\mathrm{so}}(r)
=
V^{\mathrm{so(v)}}(r)
+
V^{\mathrm{so(s)}}(r).
\end{equation}
The vector spin--orbit term is given by
\begin{equation}
V^{\mathrm{so(v)}}(r)
=
\frac{1}{r}
\frac{\mathrm{d}G(r)}{\mathrm{d}r}
\Bigg[
\frac{\mathbf{S}_1}{2m_1^2}
+
\frac{\mathbf{S}_2}{2m_2^2}
+
\frac{\mathbf{S}_1+\mathbf{S}_2}{m_1m_2}
\Bigg]
\cdot\mathbf{L},
\end{equation}
whereas the scalar spin--orbit term is
\begin{equation}
V^{\mathrm{so(s)}}(r)
=
-\frac{1}{2r}
\frac{\mathrm{d}S(r)}{\mathrm{d}r}
\left(
\frac{\mathbf{S}_1}{m_1^2}
+
\frac{\mathbf{S}_2}{m_2^2}
\right)
\cdot\mathbf{L},
\end{equation}
where $\mathbf{L}$ denotes the relative orbital angular momentum operator of the quark--antiquark pair.

Beyond the relativistic kinetic-energy terms, relativistic corrections to the quark--antiquark interaction are implemented in two steps~\cite{Godfrey:1985xj,Li:2023wgq,Song:2015nia}.
First, nonlocal effects associated with the relativistic motion of the constituent quarks are modeled by smearing the coordinate-space potentials with a mass-dependent Gaussian function.
For a radial potential $f(r)$, the smeared form is
\begin{equation}
\label{eq:smearing_general}
\tilde{f}(r)
=
\int \mathrm{d}^3\mathbf{r}^{\prime}\,
\rho_{12}(\mathbf{r}-\mathbf{r}^{\prime})f(r'),
\end{equation}
where
\begin{equation}
 \rho_{12}\left(\mathbf{r}-\mathbf{r}^{\prime}\right)
 =
 \frac{\sigma_{12}^3}{\pi^{3/2}}
 \exp\left[-\sigma_{12}^2\left(\mathbf{r}-\mathbf{r}^{\prime}\right)^2\right],
 \label{smearing}
\end{equation}
with
\begin{equation}
\begin{aligned}
\sigma_{12}^2
={}&
\sigma_0^2
\Bigg[
\frac{1}{2}
+
\frac{1}{2}
\left(
\frac{4m_1m_2}{(m_1+m_2)^2}
\right)^4
\Bigg]
\\
&+
s^2
\left(
\frac{2m_1m_2}{m_1+m_2}
\right)^2,
\end{aligned}
\label{smearing_sigma}
\end{equation}
where $\sigma_0$ and $s$ are universal smearing parameters.

Applying this prescription to $G(r)$ and $S(r)$ gives
\begin{equation}
\tilde{G}(r) = \int \mathrm{d}^3\mathbf{r}^{\prime} \rho_{12}(\mathbf{r}-\mathbf{r}^{\prime})G(r^\prime),
\end{equation}
and
\begin{equation}
\tilde{S}(r) = \int \mathrm{d}^3\mathbf{r}^{\prime}
\rho_{12} (\mathbf{r}-\mathbf{r}^{\prime})S(r^\prime).
\end{equation}
The smeared spin-dependent interactions are constructed from derivatives of $\tilde{G}(r)$ and $\tilde{S}(r)$.
We denote these terms by $\tilde{V}^{\,i}(r)$, where the index $i\in\{\mathrm{cont},\,\mathrm{tens},\,\mathrm{so(v)},\,\mathrm{so(s)}\}$ labels the contact, tensor, vector spin--orbit, and scalar spin--orbit interactions, respectively.

Second, the interactions are further relativized by introducing momentum-dependent factors that reduce to unity in the nonrelativistic limit~\cite{Godfrey:1985xj,Song:2015nia,Li:2023wgq}.
The smeared Coulomb-like potential is modified as follows:
\begin{equation}
\tilde G(r)\ \longrightarrow\
\left(1+\frac{\mathbf p^2}{E_1E_2}\right)^{\!1/2}\!
\tilde G(r)\!
\left(1+\frac{\mathbf p^2}{E_1E_2}\right)^{\!1/2},
\end{equation}
where $E_1=\sqrt{m_1^2+\mathbf{p}^2}$ and $E_2=\sqrt{m_2^2+\mathbf{p}^2}$ are the energies of the constituent quark and antiquark, respectively.

The spin-dependent interactions are also modified by momentum-dependent factors~\cite{Godfrey:1985xj,Song:2015nia}:
\begin{equation}
\label{vsoij}
\tilde V^{\,i}(r)
\longrightarrow
\left(\frac{m_1 m_2}{E_1 E_2}\right)^{\!1/2+\epsilon_i}
\tilde V^{\,i}(r)
\left(\frac{m_1 m_2}{E_1 E_2}\right)^{\!1/2+\epsilon_i},
\end{equation}
where $\epsilon_i$ is a dimensionless phenomenological parameter for interaction $i$.
The resulting interaction terms form the relativized effective potential $\tilde{V}^{\mathrm{eff}}(\mathbf{p},\mathbf{r})$ in Eq.~\eqref{eq:MGI_Hamiltonian}.

To calculate the meson spectrum, we use a simple harmonic oscillator (SHO) basis.
The basis functions in coordinate and momentum space are given by
\begin{equation}
\begin{aligned}
\Psi_{nLM_L}(\mathbf{r})
&=
R_{nL}(r,\beta)
Y_{LM_L}(\Omega_r),
\\
\Psi_{nLM_L}(\mathbf{p})
&=
R_{nL}(p,\beta)
Y_{LM_L}(\Omega_p).
\end{aligned}
\label{eq:SHO_wavefunction}
\end{equation}
The radial basis functions are
\begin{equation}
\label{eq:SHO_coordinate}
\begin{aligned}
 R_{nL}(r,\beta)=&
 N_{nL}\beta^{3/2}(\beta r)^L \\
 &\times e^{-\beta^2r^2/2}L_{n-1}^{L+\frac12}(\beta^2r^2),
\end{aligned}
\end{equation}
and
\begin{equation}
\label{eq:SHO_momentum}
\begin{aligned}
 R_{nL}(p,\beta)=&N_{nL}\frac{(-1)^{n-1}(-i)^L}{\beta^{3/2}} \\
 &\times e^{-p^2/(2\beta^2)}\left(\frac{p}{\beta}\right)^LL_{n-1}^{L+\frac12}
\left(\frac{p^2}{\beta^2}\right),
\end{aligned}
\end{equation}
with the normalization factor
\begin{equation}
N_{nL}
=
\sqrt{
\frac{2(n-1)!}
{\Gamma(n+L+\frac12)}
}.
\label{eq:normalization}
\end{equation}
Here, $n=1,2,\ldots$ labels the radial SHO basis functions, and $\beta$ is the basis scale parameter.
The function $Y_{LM_L}$ is the spherical harmonic, while $\Omega_r$ and $\Omega_p$ denote the angular coordinates of $\mathbf{r}$ and $\mathbf{p}$, respectively.
The functions $L_{n-1}^{L+\frac12}(x)$ and $\Gamma(x)$ are the generalized Laguerre polynomial and the gamma function, respectively.

The meson radial wave functions are expanded in this basis as
\begin{equation}
R_{nL}(r)
=
\sum_{k=1}^{n_{\max}}
C_k R_{kL}(r,\beta),
\label{eq:radial_r_expansion}
\end{equation}
and
\begin{equation}
R_{nL}(p)
=
\sum_{k=1}^{n_{\max}}
C_k R_{kL}(p,\beta),
\label{eq:radial_p_expansion}
\end{equation}
where $n_{\max}=21$ is the number of radial basis functions retained and $C_k$ are the expansion coefficients for the meson state under consideration.
The expansion coefficients $C_k$ are determined by diagonalizing the MGI Hamiltonian matrix in the SHO basis~\cite{Song:2015nia}.
For each state, the scale parameter $\beta$ is optimized by minimizing the corresponding eigenvalue $E_{nL}(\beta)$ with the model parameters held fixed.

\begin{table}[t]
\caption{MGI model parameters used in this work obtained in Ref.~\cite{Wang:2024lba}.}
\label{tab:parameters}
\centering
\begin{ruledtabular}
\begin{tabular}{cc@{\qquad}cc}
Parameter & Value & Parameter & Value \\
\hline
$m_{u,d}$ (GeV) & 0.162 & $m_s$ (GeV) & 0.377 \\
$b$ (GeV$^2$) & 0.222 & $c$ (GeV) & $-0.228$ \\
$\mu$ (GeV) & 0.0779 & $\sigma_0$ (GeV) & 1.79 \\
$s$ & 0.711 & $\epsilon_{\mathrm{cont}}$ & $-0.137$ \\
$\epsilon_{\mathrm so(v)}$ & 0.0550 & $\epsilon_{\mathrm so(s)}$ & 0.366 \\
$\epsilon_{\mathrm{tens}}$ & 0.493 & & \\
\end{tabular}
\end{ruledtabular}
\end{table}

\subsection{The QPC model}

The QPC model was first proposed by Micu~\cite{Micu:1968mk} and later developed by the Orsay group~\cite{LeYaouanc:1972vsx,LeYaouanc:1973ldf,LeYaouanc:1974cvx}.
It has been widely used to study OZI-allowed two-body strong decays of mesons~\cite{Song:2015fha,Chen:2026mdo,Chen:2015iqa,Pang:2018gcn,Bonnaz:2001aj,Zhou:2004mw,Lu:2006ry,Zhang:2006yj,Luo:2009wu,Sun:2009tg,Liu:2009fe,Sun:2010pg,Rijken:2010zza,Ye:2012gu,Wang:2012wa,He:2013ttg,Sun:2013qca,Wang:2020due}.
In this framework, meson decay proceeds through the creation of a quark--antiquark pair from the vacuum with quantum numbers $J^{PC}=0^{++}$.
For the two-body decay $A\to B+C$, the transition operator $\mathcal{T}$ is given by~\cite{Pang:2018gcn,Chen:2026mdo}
\begin{align}
\mathcal{T}={}&-3\gamma\sum_m\langle1m;1,-m|00\rangle
\int\dd^3\mathbf{p}_3\dd^3\mathbf{p}_4\,
\delta^{(3)}(\mathbf{p}_3+\mathbf{p}_4)\nonumber\\
&\times\mathcal Y_{1m}\left(\frac{\mathbf{p}_3-\mathbf{p}_4}{2}\right)
\chi_{1,-m}^{34}\phi_0^{34}\omega_0^{34}
b_{3i}^\dagger(\mathbf{p}_3)d_{4j}^\dagger(\mathbf{p}_4),
\label{eq:qpc}
\end{align}
where $\gamma$ is the dimensionless pair-creation strength.
In this work, the same pair-creation strength $\gamma$ is used for the creation of $u\bar u$, $d\bar d$, and $s\bar s$ pairs.
Its value is determined from the measured total width of $f_4(2050)$.
The operators $b_{3i}^\dagger(\mathbf{p}_3)$ and $d_{4j}^\dagger(\mathbf{p}_4)$ create the quark and antiquark labeled $3$ and $4$, with three-momenta $\mathbf{p}_3$ and $\mathbf{p}_4$, respectively.
The color indices $i,j=1,2,3$ are summed over.
The functions $\chi_{1,-m}^{34}$, $\phi_0^{34}$, and $\omega_0^{34}$ are the spin, flavor, and color wave functions of the created pair, respectively.
The solid harmonic is defined as $\mathcal{Y}_{\ell m}(\mathbf{p})\equiv |\mathbf{p}|^\ell Y_{\ell m}(\theta_p,\phi_p)$, where $\ell$ denotes the orbital angular momentum quantum number.

In the rest frame of the initial meson $A$, the decay amplitude is defined by
\begin{equation}
\langle BC|\mathcal{T}|A\rangle = \delta^{(3)}(\mathbf{P}_B + \mathbf{P}_C)\mathcal{M}^{M_{J_A}M_{J_B}M_{J_C}},
\end{equation}
where $\mathbf{P}_B$ and $\mathbf{P}_C$ are the three-momenta of the final-state mesons, and $M_{J_X}$ denotes the projection of the total angular momentum $J_X$ of meson $X$ ($X=A,B,C$).

We define $\mathbf{P}=\mathbf{P}_B=-\mathbf{P}_C$.
The two-body partial width takes the form
\begin{equation}
\Gamma_{A\to BC}
=
\frac{\pi |\mathbf{P}|}{4M_A^2}
\sum_{J,L}
\left|\mathcal{M}^{JL}(\mathbf{P})\right|^2,
\label{eq:qpc_width}
\end{equation}
where $M_A$ is the mass of meson $A$, $L$ is the relative orbital angular momentum between the final-state mesons, and $J$ is the total angular momentum obtained by coupling $J_B$ and $J_C$.

The partial-wave amplitude is related to the decay amplitude through the Jacob--Wick transformation~\cite{Jacob:1959at}:
\begin{equation}
\begin{aligned}
\mathcal{M}^{JL}(\mathbf{P})
={}&
\frac{\sqrt{4\pi(2L+1)}}{2J_A+1}
\sum_{M_{J_A},M_{J_B},M_{J_C}}
\\
&\times
\langle L0;JM_{J_A}|J_A M_{J_A}\rangle
\\
&\times
\langle J_B M_{J_B};J_C M_{J_C}|JM_{J_A}\rangle
\\
&\times
\mathcal{M}^{M_{J_A}M_{J_B}M_{J_C}}
(\mathbf{P}).
\end{aligned}
\label{eq:jacob_wick}
\end{equation}
The decay amplitude entering this transformation can be written as
\begin{equation}
\begin{aligned}
&\mathcal{M}^{M_{J_A}M_{J_B}M_{J_C}}(\mathbf{P})
\\
&=
\gamma\sqrt{8E_AE_BE_C}
\sum_{\substack{
M_{L_A},M_{S_A}\\
M_{L_B},M_{S_B}\\
M_{L_C},M_{S_C},m}}
\\
&\quad\times
\langle L_A M_{L_A} S_A M_{S_A}
|J_A M_{J_A}\rangle
\\
&\quad\times
\langle L_B M_{L_B} S_B M_{S_B}
|J_B M_{J_B}\rangle
\\
&\quad\times
\langle L_C M_{L_C} S_C M_{S_C}
|J_C M_{J_C}\rangle
\\
&\quad\times
\langle 1m;1,-m|00\rangle
\\
&\quad\times
\langle
\chi^{14}_{S_B M_{S_B}}
\chi^{32}_{S_C M_{S_C}}
|
\chi^{12}_{S_A M_{S_A}}
\chi^{34}_{1,-m}
\rangle
\\
&\quad\times
\Big[
\langle
\phi_B^{14}\phi_C^{32}
|
\phi_A^{12}\phi_0^{34}
\rangle
I(\mathbf{P};m_1,m_2,m_3)
\\
&\qquad+
(-1)^{1+S_A+S_B+S_C}
\langle
\phi_B^{32}\phi_C^{14}
|
\phi_A^{12}\phi_0^{34}
\rangle
\\
&\qquad\quad\times
I(-\mathbf{P};m_2,m_1,m_3)
\Big].
\end{aligned}
\label{eq:qpc_amplitude}
\end{equation}
Here, $E_A=M_A$, $E_B=\sqrt{M_B^2+|\mathbf{P}|^2}$, and $E_C=\sqrt{M_C^2+|\mathbf{P}|^2}$ are the meson energies.
$L_X$ and $S_X$ denote the internal orbital angular momentum and total quark spin of meson $X$, respectively, and $M_{L_X}$ and $M_{S_X}$ are their projections.

The spatial overlap integral is
\begin{equation}
\begin{aligned}
&I(\mathbf{P};m_1,m_2,m_3)
\\
&=
\int \mathrm{d}^3\mathbf{k}\,
\psi_{n_B L_B M_{L_B}}^*(\mathbf{k}+U\mathbf{P})
\\
&\quad\times
\psi_{n_C L_C M_{L_C}}^*(\mathbf{k}+V\mathbf{P})
\\
&\quad\times
\psi_{n_A L_A M_{L_A}}(\mathbf{k}+\mathbf{P})
\mathcal{Y}_{1m}(\mathbf{k}),
\end{aligned}
\label{eq:overlap_integral}
\end{equation}
where $\mathbf{k}=\mathbf{p}_3$, and
\begin{equation}
U=\frac{m_3}{m_1+m_3},
\qquad
V=\frac{m_3}{m_2+m_3}.
\label{eq:UV}
\end{equation}
The masses $m_1$ and $m_2$ refer to the constituent quark and antiquark in meson $A$, while $m_3$ is the mass of the created quark, with $m_4=m_3$.
The functions $\psi_{n_X L_X M_{L_X}}$ are the momentum-space wave functions of the mesons, with $n_X$ denoting the radial quantum number.
The dependence of $I$ on $M_{L_A}$, $M_{L_B}$, $M_{L_C}$, and $m$ is implicit, with $m=-1,0,1$.

The magnitude of the final-state momentum is
\begin{equation}
|\mathbf{P}|=
\frac{
\sqrt{
[M_A^2-(M_B+M_C)^2]
[M_A^2-(M_B-M_C)^2]
}
}{2M_A},
\label{eq:momentum}
\end{equation}
where $M_B$ and $M_C$ are the masses of the final-state mesons.

The two types of mixing schemes include spin mixing and flavor mixing.
The spin mixing scheme for the kaons can be expressed as
\begin{equation}\label{anglek1}
\left( \begin{matrix}
 |K(nL)\rangle \\
 |K^\prime(nL)\rangle \\
\end{matrix}\right) =
\left( \begin{matrix}
 \textrm{$\cos\theta_{nL}$} & \textrm{$\sin\theta_{nL}$} \\
 \textrm{$-\sin\theta_{nL}$} & \textrm{$\cos\theta_{nL}$} \\
\end{matrix}\right)
\left( \begin{matrix}
 |K(n^1L_L)\rangle \\
 |K(n^3L_L)\rangle \\
\end{matrix}\right),
\end{equation}
where $\theta_{nL}$ is the mixing angle between the $K(n^1L_L)$ and the $K(n^3L_L)$ states.

The flavor mixing takes the form
\begin{equation}\label{mixingns}
\left( \begin{matrix}
 X \\
 X^\prime \\
\end{matrix}\right) =
\left( \begin{matrix}
 \textrm{$\cos\theta_{x}$} & \textrm{$\sin\theta_{x}$} \\
 \textrm{$-\sin\theta_{x}$} & \textrm{$\cos\theta_{x}$} \\
\end{matrix}\right)
\left( \begin{matrix}
 |n\bar{n}\rangle \\
 |s\bar{s}\rangle \\
\end{matrix}\right),
\end{equation}
where $X$ and $X^\prime$ are the two isoscalar mesons (such as $\eta$ and $\eta^\prime$), and $\theta_x$ is the mixing angle in the quark flavor scheme.
\par
In this work, the mixing angles adopted for the mesons are listed in Table~\ref{tab:decay_mixing_angles}.

For each $n\,{}^3L_4$ configuration, we describe the mixing of the two isoscalar states in the quark-flavor basis, with $|n\bar n\rangle$ and $|s\bar s\rangle$.
The states are written as
\begin{equation}
\begin{pmatrix}
|f_4\rangle\\
|f'_4\rangle
\end{pmatrix}
=
\begin{pmatrix}
\cos\theta_{n\,{}^3L_4}
&
\sin\theta_{n\,{}^3L_4}
\\
-\sin\theta_{n\,{}^3L_4}
&
\cos\theta_{n\,{}^3L_4}
\end{pmatrix}
\begin{pmatrix}
|n\bar n\rangle\\
|s\bar s\rangle
\end{pmatrix}.
\label{eq:f4_flavor_mixing}
\end{equation}
Here, $\theta_{n\,{}^3L_4}$ is the flavor-mixing angle for the specified configuration.
The $n$ in this spectroscopic label denotes the radial quantum number, while $L=3$ or $5$ specifies the internal orbital angular momentum of the quark--antiquark state.
\begin{table}[tbp]
\caption{Mixing angles used in the strong-decay calculations.
The angles $\theta_{nL}$ describe mixing between the $n\,{}^1L_L$ and $n\,{}^3L_L$ configurations, while $\theta_{n\,{}^{2S+1}L_J}$ describe nonstrange--strange flavor mixing.
All angles are given in degrees.}
\label{tab:decay_mixing_angles}
\centering
\small
\renewcommand{\arraystretch}{1}
\begin{ruledtabular}
\begin{tabular}{cccc}
Angle & $n$ & Value (deg) & Reference \\
\hline
$\theta_{nP}$ & $1$ & $-34.0$
& \cite{Cheng:2011pb} \\
$\theta_{nP}$ & $2$ & $-14.0$
& \cite{Li:2024hrf} \\
$\theta_{nP}$ & $3$ & $-12.8$
& \cite{Li:2024hrf} \\
$\theta_{nD}$ & $1,2,3$ & $-39.0$\textsuperscript{a}
& \cite{Pang:2017dlw} \\
$\theta_{nF}$ & $1,2$ & $-40.9$\textsuperscript{a}
& \cite{Bokade:2025lmn} \\
$\theta_{nG}$ & $1$ & $-41.8$\textsuperscript{a}
& \cite{Bokade:2025lmn} \\
\hline
$\theta_{n\,{}^{1}S_0}$ & $1$ & $-39.3$
& \cite{Feldmann:1999uf} \\
$\theta_{n\,{}^{1}S_0}$ & $2,3,4$ & $-39.3$\textsuperscript{b}
&\cite{Yu:2011ta}\\
$\theta_{n\,{}^{3}S_1}$ & $1,2,3$ & $0$
& \textemdash \\
$\theta_{n\,{}^{3}P_0}$ & $1,2$ & $0$
& \textemdash \\
$\theta_{n\,{}^{1}P_1}$ & $1$ & $4.4$
& \cite{Li:2005eq,Chen:2015iqa} \\
$\theta_{n\,{}^{1}P_1}$ & $2,3$ & $4.4$\textsuperscript{b}
& \cite{Chen:2015iqa,Wang:2024lba} \\
$\theta_{n\,{}^{3}P_1}$ & $1$ & $-24.0$\textsuperscript{c}
& \cite{LHCb:2013ged} \\
$\theta_{n\,{}^{3}P_1}$ & $2,3,4$ & $-24.0$\textsuperscript{b}
& \textemdash \\
$\theta_{n\,{}^{3}P_2}$ & $1$ & $0$
& \cite{Geng:2010kma} \\
$\theta_{n\,{}^{3}P_2}$ & $2,3$ & $0$
& \textemdash \\
$\theta_{n\,{}^{1}D_2}$ & $1,2,3$ & $0$
& \textemdash \\
$\theta_{n\,{}^{3}D_1}$ & $1,2$ & $0$
& \textemdash \\
$\theta_{n\,{}^{3}D_2}$ & $1,2$ & $0$
& \textemdash \\
$\theta_{n\,{}^{3}D_3}$ & $1,2$ & $0$
& \textemdash \\
$\theta_{n\,{}^{1}F_3}$ & $1$ & $0$
& \textemdash \\
$\theta_{n\,{}^{3}F_2}$ & $1,2$ & $0$
& \textemdash \\
$\theta_{n\,{}^{3}F_3}$ & $1,2$ & $0$
& \textemdash \\
\hline
$\theta_{n\,{}^{3}F_4}$ & $1$ & $6$\textsuperscript{d}
& \cite{Dudek:2013yja} \\
$\theta_{n\,{}^{3}F_4}$ & $2,3,4$ & $6$\textsuperscript{b}
& \textemdash \\
$\theta_{n\,{}^{3}H_4}$ & $1,2$ & $0$
& \textemdash \\
\end{tabular}
\end{ruledtabular}

\par\smallskip
{\footnotesize\raggedright Zero flavor-mixing angles correspond to the ideal-mixing approximation, with pure $n\bar n$ and $s\bar s$ states.
\par
\textsuperscript{a} The magnitudes of these angles follow the heavy-quark-limit expression $|\theta_{nL}|\simeq\arctan\sqrt{L/(L+1)}$, used here as a reference for the kaon states.
\par
\textsuperscript{b} The ground-state angle is also used for the listed radial excitations.
\par
\textsuperscript{c} The measurement determines the magnitude of the angle; the negative branch is adopted here.
\par
\textsuperscript{d} The interval $(6\pm4)^\circ$ is adopted here, with central values evaluated at $6^\circ$.

\par}

\end{table}

\begin{table}[htbp]
\centering
\caption{Mass spectrum of the $f_4$ family, in MeV.
The MGI and GI results are calculated in this work.}
\label{tab:f4mass}
\renewcommand{\arraystretch}{1}
\begin{ruledtabular}
\begin{tabular}{ccccccc}
Flavor
& $n\,{}^{2S+1}L_J$& State & MGI & Expt.\textsuperscript{b} & GI & Ebert\textsuperscript{c}\\
\hline
\multirow{6}{*}{$n\bar n$}
& $1^3F_4$& $f_4(2050)$&1973&$2018\pm11$&2008&2018\\
& $2^3F_4$&$f_4(2300)$\textsuperscript{a} &2243&$2320\pm60$ &2409&2284\\
& $3^3F_4$&&2466&\textemdash&2775&\textemdash\\
& $4^3F_4$&&2640&\textemdash&3114&\textemdash\\
& $1^3H_4$&&2406&\textemdash&2644&2234\\
& $2^3H_4$&&2589&\textemdash&2971&\textemdash\\
\hline
\multirow{6}{*}{$s\bar s$}
& $1^3F_4$&$f_4(2300)$\textsuperscript{a} &2170&$2320\pm60$&2202& 2286 \\
& $2^3F_4$&&2453&\textemdash&2597&2657\\
& $3^3F_4$&&2688&\textemdash&2953&\textemdash\\
& $4^3F_4$&&2878&\textemdash&3282&\textemdash\\
& $1^3H_4$&&2607&\textemdash&2803&2634\\
& $2^3H_4$&&2802&\textemdash&3121&\textemdash\\
\end{tabular}
\end{ruledtabular}

\par\smallskip
{\footnotesize\raggedright

\textsuperscript{a} The two entries for $f_4(2300)$ represent alternative assignments of the same resonance.
\par
\textsuperscript{b} Experimental masses are taken from the PDG~\cite{ParticleDataGroup:2026mpi}.
The value for $f_4(2300)$ is the PDG estimate.
\par
\textsuperscript{c} Results from Ref.~\cite{Ebert:2009ub}.

\par}
\end{table}
\section{NUMERICAL RESULTS AND PHENOMENOLOGICAL ANALYSIS}\label{sec3}
We calculate the masses of the $f_4$ states using the MGI model, as listed in Table~\ref{tab:f4mass}, and their OZI-allowed two-body strong decays using the QPC model.
We now turn to a phenomenological analysis of the spectrum and the decay information of the $f_4$ light meson family.

\subsection{Mass spectra analysis of $f_4$ meson family}

The experimental masses, GI predictions, and results of Ebert \textit{et al.}~\cite{Ebert:2009ub} are included for comparison in Table~\ref{tab:f4mass}.
In the MGI spectrum, the radial level spacings decrease with increasing excitation in both flavor sectors.
For the nonstrange $F$-wave sequence, the spacing decreases from $270\,\mathrm{MeV}$ between $1F$ and $2F$ to $174\,\mathrm{MeV}$ between $3F$ and $4F$.
The MGI masses are lower than the corresponding GI values, with larger differences for higher excitations.
The differences are $35\,\mathrm{MeV}$ for the nonstrange $1\,{}^3F_4$ state and $474\,\mathrm{MeV}$ for the $4\,{}^3F_4$ state.

We adopt the conventional nonstrange $1\,{}^3F_4$ assignment for $f_4(2050)$.
For $f_4(2300)$, the MGI masses of the $2\,{}^3F_4(n\bar n)$ and $1\,{}^3F_4(s\bar s)$ configurations are 2243 and $2170\,\mathrm{MeV}$, respectively.
These values are 77 and $150\,\mathrm{MeV}$ below the central PDG estimate of $2320\pm60\,\mathrm{MeV}$~\cite{ParticleDataGroup:2026mpi}.
The nonstrange radial excitation is therefore closer to the observed mass in the MGI spectrum.
However, Ebert \textit{et al.} predict nearly degenerate masses of 2284 and $2286\,\mathrm{MeV}$ for these two configurations~\cite{Ebert:2009ub}.

The MGI mass of the nonstrange $1\,{}^3H_4$ state is $2406\,\mathrm{MeV}$, which lies $86\,\mathrm{MeV}$ above the central mass of $f_4(2300)$.
Its mass offset is comparable to that of the nonstrange $2\,{}^3F_4$ state, so mass proximity alone does not exclude this configuration.
These comparisons motivate the strong-decay analysis below, which provides complementary information for distinguishing the flavor and orbital assignments.
The calculated decay properties are presented in Tables~\ref{tab:channels24}--\ref{tab:channels_1H_2H_theta0} and Figures~\ref{fig:pair1F} and \ref{fig:34F12H}.

\subsection{Strong decay analysis of the $f_4$ meson family}
\subsubsection{The ground state of the $f_4$ family: $f_4(2050)$}

\begin{table}[tbp]
\caption{Two-body strong decays of $f_4(2050)$ at $M=2018\,\mathrm{MeV}$ and $\theta_{1\,{}^3F_4}=6^\circ$.
Widths are in MeV.
Charge multiplicities and charge-conjugate modes are included.}
\label{tab:channels24}
\centering
\footnotesize\setlength{\tabcolsep}{1.25pt}
\begin{ruledtabular}
\begin{tabular}{cccccc}
Mode
& $\Gamma_{\mathrm{th}}$ [range]\textsuperscript{a}
& $\mathcal B_{\mathrm{th}}$ (\%) & $\Gamma_{\mathrm{B}}$\textsuperscript{d}~\cite{Barnes:1996ff}
& $\mathcal B_{\mathrm{exp}}$ (\%)\textsuperscript{b}
& $\Gamma_{\mathrm{der}}$\textsuperscript{c} \\
\hline
$\rho\,\rho$
& {$100$ $[97, 110]$}
& {$44$}
& $86$
& \textemdash
& \textemdash \\
$a_{2}\,\pi$
& {$40$ $[37, 44]$}
& {$17$}
& $22$
& seen
& \textemdash \\
$\pi\,\pi$
& {$30$ $[28, 33]$}
& {$13$}
& $62$
& $17\pm1.5$
& $40$ \\
$\omega\,\omega$
& {$26$ $[24, 27]$}
& {$11$}
& $27$
& seen
& ${60}$ \\
$a_{1}\,\pi$
& {$14$ $[13, 15]$}
& {$5.9$}
& $9$
& \textemdash
& \textemdash \\
$\pi\,\pi(1300)$
& {$9.4$ $[8.7, 10]$}
& {$4.0$}
& $2$
& \textemdash
& \textemdash \\
$\pi_{2}\,\pi$
& {$3.2$ $[3.0, 3.5]$}
& {$1.4$ }
& $1$
& \textemdash
& \textemdash \\
$K^{*}\,\bar K^{*}$
& {$4.1$ $[3.8, 4.5]$}
& {$1.8$}
& $10$
& \textemdash
& \textemdash \\
$f_{2}\,\eta$
& {$1.3$ $[1.2, 1.3]$}
& {$0.53$}
& $1$
& \textemdash
& \textemdash \\
$\eta\,\eta$
& {$1.2$ $[1.1, 1.3]$}
& {$0.51$}
& $2$
& $0.21\pm0.08$
& 0.50 \\
$K\,\bar K$
& {$1.0$ $[0.96, 1.1]$}
& {$0.44$}
& $9$
& $0.68^{+0.34}_{-0.18}$
& 1.6 \\
$a_{1}(1640)\,\pi$
& {$0.35$ $[0.33, 0.38]$}
& {$0.15$}
& $0$
& \textemdash
& \textemdash \\
$\eta'\,\eta$
& {$0.22$ $[0.20, 0.23]$}
& {$0.089$}
& $0$
& \textemdash
& \textemdash \\
$h_{1}\,\omega$
& {$0.22$ $[0.20, 0.24]$}
& {$0.092$}
& \textemdash
& \textemdash
& \textemdash \\
$f_{1}\,\eta$
& {$0.22$ $[0.20, 0.24]$}
& {$0.093$}
& $0$
& \textemdash
& \textemdash \\
$K\,\bar K^{*}$
& {$0.058$ $[0.054, 0.063]$}
& {$0.025$ }
& $5$
& \textemdash
& \textemdash \\
Others
& {$0.36$ $[0.33, 0.38]$}
& {$0.15$}
& \textemdash
& \textemdash
& \textemdash \\
\hline
Total
& {$237$ $[219, 255]$}
& 100
& $237$
& 100
& $237\pm18$ \\
\end{tabular}
\end{ruledtabular}

\par\smallskip
{\footnotesize\raggedright

\textsuperscript{a} Central widths are evaluated at $\gamma=6.90$; brackets give the range for $\gamma=6.64$--$7.16$, with all other inputs fixed.
Branching fractions are normalized to the sum over all calculated channels and are independent of $\gamma$.
Other channels collects the remaining calculated modes.
\par
\textsuperscript{b} Experimental information is taken from Ref.~\cite{ParticleDataGroup:2026mpi}.
``seen'' denotes an observed mode without a quoted branching fraction; a dash denotes no entry in the cited PDG listing.
\par
\textsuperscript{c} Derived partial widths are obtained from $\Gamma_{\mathrm{der}} =\mathcal B_{\mathrm{exp}}\times237\,\mathrm{MeV}$; only central values are quoted.
The total-width entry is the PDG value.
\par
$^{\dagger}$This value additionally uses $\Gamma_{\omega\omega}/\Gamma_{\pi\pi}=1.5\pm0.3$.
\par
\textsuperscript{d}$\Gamma_{\mathrm{B}}$ gives the \({}^3P_0\) predictions for $f_4(2044)$ from Table B12 of Ref.~\cite{Barnes:1996ff}.
\par}

\end{table}

We take $f_4(2050)$ as the mainly nonstrange $1\,{}^3F_4$ reference state for the decay analysis of the family.
Its measured total width determines the common QPC pair-creation strength, while the relative decay rates provide further tests of the calculation.

The decay amplitudes are evaluated using the numerical MGI wave functions and the experimental parent mass $M=2018\,\mathrm{MeV}$.
We adopt the reference flavor-mixing angle $\theta_{1\,{}^3F_4}=6^\circ$ specified in Table~\ref{tab:decay_mixing_angles}.
Matching the sum of the calculated two-body strong-decay widths to the experimental total width, $237\pm18\,\mathrm{MeV}$~\cite{ParticleDataGroup:2026mpi}, gives $\gamma=6.90^{+0.26}_{-0.27}$.
The uncertainty is obtained by propagating the experimental total-width uncertainty, with all other inputs held fixed.

The partial widths and branching fractions are listed in Table~\ref{tab:channels24}.
The four leading channels are $\rho\rho$, $a_2(1320)\pi$, $\pi\pi$, and $\omega\omega$, which together account for approximately $85\%$ of the calculated total width.
The largest contribution is $\rho\rho$, with a branching fraction of $44\%$.
The experimentally observed $a_2(1320)\pi$ and $\omega\omega$ modes also have sizeable predicted branching fractions of $17\%$ and $11\%$, respectively.

Barnes \textit{et al.} also found $\rho\rho$ to be the largest channel in a ${}^3P_0$ calculation using SHO wave functions and a parent mass of $2044\,\mathrm{MeV}$~\cite{Barnes:1996ff}.
They obtained partial widths of $62\,\mathrm{MeV}$ for $\pi\pi$ and $22\,\mathrm{MeV}$ for $a_2\pi$, compared with $31$ and $41\,\mathrm{MeV}$, respectively, in the present calculation.
The ordering of these two channels is therefore reversed.
The predicted $\omega\omega$ widths are similar: $27\,\mathrm{MeV}$ in their calculation and $26\,\mathrm{MeV}$ in ours.

The comparison with experimental branching fractions reveals quantitative differences.
The predicted $\pi\pi$ and $K\bar K$ fractions, $13\%$ and $0.44\%$, are close to the PDG values of $(17.0\pm1.5)\%$ and $(0.68^{+0.34}_{-0.18})\%$, respectively.
The predicted $\eta\eta$ fraction, $0.51\%$, is comparable with the PDG value of $(0.21\pm0.08)\%$~\cite{ParticleDataGroup:2026mpi}.
We obtain $\Gamma_{\omega\omega}/\Gamma_{\pi\pi}=0.84$ and $\Gamma_{K\bar K}/\Gamma_{\pi\pi}=0.034$, which are comparable with the measured values of $1.5\pm0.3$ and $0.04^{+0.02}_{-0.01}$, respectively~\cite{ParticleDataGroup:2026mpi}.
The $\omega\omega/\pi\pi$ ratio lies slightly below its experimental interval, while the $K\bar K/\pi\pi$ ratio lies within its quoted interval.

A measurement of the $\rho\rho$ branching fraction would test the predicted dominance of this channel.
We use the calibrated pair-creation strength and its fitted range in the following analysis of the excited states and strange partners.

\subsubsection{$F$-wave assignments of $f_4(2300)$ and predictions for $f'_4(2F)$}
\begin{table*}[tbp]
\caption{Two-body strong-decay widths and branching fractions for the $f_4(2F)$ and $f'_4(1F)$ assignments of $f_4(2300)$ and for the predicted $f'_4(2F)$ state.
The calculations use $\gamma=6.90$ and $\theta_{1\,{}^3F_4}=\theta_{2\,{}^3F_4}=6^\circ$.
The parent mass is set to 2320 MeV for both assignments of $f_4(2300)$ and to 2453.23 MeV for $f'_4(2F)$.
Widths are in MeV.
Charge-conjugate modes and all allowed charge combinations are included.
Branching fractions are normalized to the total calculated two-body width.
Channels with $\mathcal B_i\geq2\%$ in any of the three cases are listed explicitly; the remaining channels are grouped under Other channels.
Dashes denote entries for which no numerical value is quoted, including partial widths below $10^{-3}$ MeV.
Calculated widths and branching fractions are rounded to three significant figures.}
\label{tab:f42300assignments_theta3}
\centering
\footnotesize\setlength{\tabcolsep}{3pt}
\begin{ruledtabular}
\begin{tabular}{lrrrrrr}
 & \multicolumn{2}{c}{$f_4(2300)$ as $f_4(2F)$} & \multicolumn{2}{c}{$f_4(2300)$ as $f'_4(1F)$} & \multicolumn{2}{c}{Prediction: $f'_4(2F)$} \\
Mode & $\Gamma_i$ & $\mathcal B_i$ (\%) & $\Gamma_i$ & $\mathcal B_i$ (\%) & $\Gamma_i$ & $\mathcal B_i$ (\%) \\
\hline
$a_{2}(1700)\,\pi$ & $38.7$ & $16.7$ & $0.347$ & $0.0923$ & $0.954$ & $0.441$ \\
$\rho\,\rho$ & $27.4$ & $11.8$ & $1.59$ & $0.424$ & $0.573$ & $0.265$ \\
$\pi\,\pi(1300)$ & $17.1$ & $7.39$ & $0.998$ & $0.266$ & $0.194$ & $0.0898$ \\
$a_{1}(1640)\,\pi$ & $16.5$ & $7.11$ & $0.131$ & $0.0348$ & $0.319$ & $0.147$ \\
$\pi_{2}(1880)\,\pi$ & $15.9$ & $6.88$ & $0.0190$ & $0.00506$ & $0.353$ & $0.163$ \\
$b_{1}(1235)\,\rho$ & $12.1$ & $5.23$ & $1.10$ & $0.293$ & $0.192$ & $0.0888$ \\
$\omega\,\omega$ & $8.86$ & $3.82$ & $0.522$ & $0.139$ & $0.141$ & $0.0652$ \\
$\pi_{2}(1670)\,\pi$ & $7.56$ & $3.26$ & $0.491$ & $0.131$ & $0.120$ & $0.0553$ \\
$a_{2}(1320)\,\pi$ & $7.28$ & $3.14$ & $2.59$ & $0.689$ & $0.178$ & $0.0822$ \\
$\omega\,\omega(1420)$ & $10.7$ & $4.61$ & $0.0805$ & $0.0214$ & $0.275$ & $0.127$ \\
$\rho\,\rho(1450)$ & $17.9$ & $7.75$ & $0.129$ & $0.0344$ & $0.758$ & $0.350$ \\
$\pi\,\pi$ & $5.90$ & $2.55$ & $0.572$ & $0.152$ & $0.178$ & $0.0821$ \\
$a_{3}(1875)\,\pi$ & $4.90$ & $2.12$ & $0.505$ & $0.135$ & $0.0604$ & $0.0279$ \\
$a_{4}(2040)\,\pi$ & $6.14$ & $2.65$ & $0.427$ & $0.114$ & $0.144$ & $0.0667$ \\
$h_{1}(1170)\,\omega$ & $4.76$ & $2.05$ & $0.556$ & $0.148$ & $0.0689$ & $0.0318$ \\
$a_{3}(2030)\,\pi$ & $7.47$ & $3.22$ & $0.121$ & $0.0323$ & $0.484$ & $0.224$ \\
$K^{*}\,\bar K^{*}$ & $0.689$ & $0.297$ & $102$ & $27.2$ & $39.4$ & $18.2$ \\
$K\,\bar K^{*}$ & $0.202$ & $0.0874$ & $68.8$ & $18.3$ & $10.4$ & $4.83$ \\
$K\,\bar K_{1}(1270)$ & $0.0779$ & $0.0336$ & $43.2$ & $11.5$ & $2.30$ & $1.06$ \\
$K\,\bar K$ & $0.0348$ & $0.0150$ & $22.9$ & $6.10$ & $9.68$ & $4.47$ \\
$K\,\bar K_{2}^{*}(1430)$ & $0.645$ & $0.278$ & $43.0$ & $11.4$ & $4.68$ & $2.16$ \\
$\phi\,\phi$ & $0.00232$ & $0.00100$ & $27.1$ & $7.20$ & $1.94$ & $0.895$ \\
$K^{*}\,\bar K^{*}(1410)$ & $0.0477$ & $0.0206$ & $0.430$ & $0.114$ & $59.0$ & $27.3$ \\
$K\,\bar K^{*}(1410)$ & $0.136$ & $0.0587$ & $21.0$ & $5.59$ & $19.3$ & $8.91$ \\
$K^{*}\,\bar K_{2}^{*}(1430)$ & $0.00164$ & $7.10\times10^{-4}$ & $3.08$ & $0.820$ & $10.3$ & $4.77$ \\
$K\,\bar K(1460)$ & $0.406$ & $0.175$ & $10.3$ & $2.76$ & $8.80$ & $4.07$ \\
$K_{1}(1270)\,\bar K^{*}$ & $0.268$ & $0.116$ & $5.71$ & $1.52$ & $7.88$ & $3.64$ \\
$K\,\bar K_{2}(1770)$ & $0.0257$ & $0.0111$ & $0.0719$ & $0.0192$ & $7.61$ & $3.52$ \\
$K\,\bar K_{1}(1790)$ & \textemdash & \textemdash & $0.0164$ & $0.00436$ & $8.17$ & $3.77$ \\
$K\,\bar K_{3}^{*}(1780)$ & $0.0200$ & $0.00862$ & $1.81$ & $0.482$ & $6.32$ & $2.92$ \\
$K\,\bar K_{1}(1400)$ & $0.319$ & $0.138$ & $5.14$ & $1.37$ & $5.66$ & $2.61$ \\
$a_{2}(3P)\,\pi$ & $5.12$ & $2.21$ & $0.00665$ & $0.00177$ & $0.349$ & $0.161$ \\
\hline
Other channels & $14.4$ & $6.23$ & $10.5$ & $2.80$ & $9.61$ & $4.44$ \\
Total & $232$ & $100$ & $375$ & $100$ & $216$ & $100$ \\
\end{tabular}
\end{ruledtabular}
\end{table*}

\begin{figure}[tbp]
\includegraphics[width=\columnwidth]{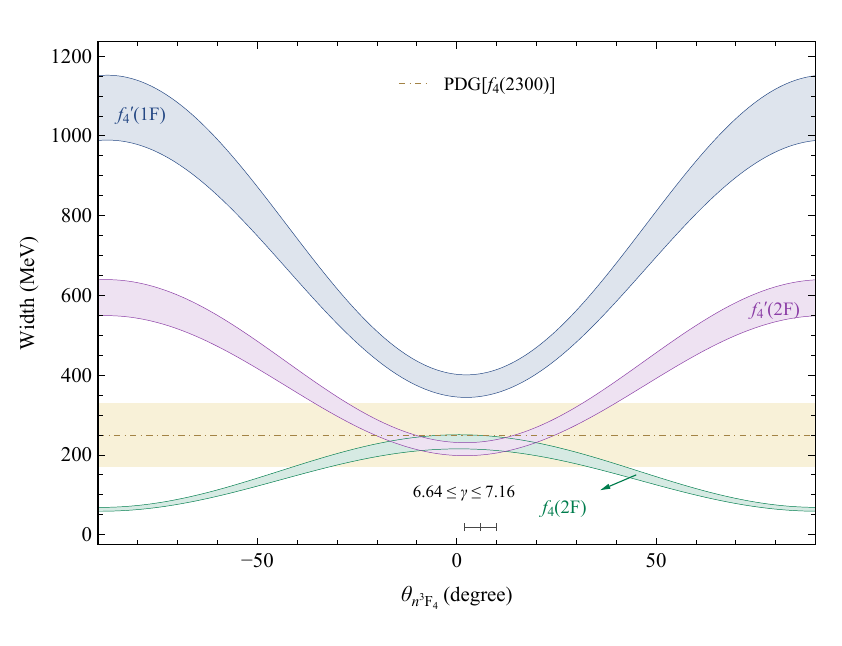}
\caption{Total widths of $f'_4(1F)$, $f_4(2F)$, and $f'_4(2F)$.
The horizontal axis represents $\theta_{1\,{}^3F_4}$ for $f'_4(1F)$ and $\theta_{2\,{}^3F_4}$ for the two $2F$ states.
The theoretical bands cover $6.64\leq\gamma\leq7.16$ with all other inputs fixed.
The parent masses are $2320$, $2320$, and $2453$ MeV, respectively.
The pale yellow band shows the PDG width of $250\pm80$ MeV for $f_4(2300)$~\cite{ParticleDataGroup:2026mpi}, with the dash-dotted line marking the central value.
The bracket marks the adopted $F$-wave interval from $2^\circ$ to $10^\circ$, with the middle tick at $6^\circ$.}
\label{fig:pair1F}
\end{figure}

We examine two $F$-wave assignments of $f_4(2300)$: the mainly nonstrange radial excitation $f_4(2F)$ and the mainly strange ground state $f'_4(1F)$.
Their OZI-allowed two-body strong decays are calculated using the QPC model with the MGI spatial wave functions.
We use the common pair-creation strength $\gamma=6.90$ determined from $f_4(2050)$ and the reference mixing angles $\theta_{1\,{}^3F_4}=\theta_{2\,{}^3F_4}=6^\circ$.
The results are summarized in Table~\ref{tab:f42300assignments_theta3} and Figure~\ref{fig:pair1F}.

For the $f_4(2F)$ assignment, the calculated total width is 232 MeV at an input mass of 2320 MeV.
The largest partial widths are those of $a_2(1700)\pi$ and $\rho\rho$, at 38.7 and 27.4 MeV, corresponding to branching fractions of 16.7\% and 11.8\%, respectively.
The $\pi\pi(1300)$, $a_1(1640)\pi$, $\pi_2(1880)\pi$, and $\rho\rho(1450)$ channels also make sizable contributions.
Together, these six channels account for approximately $57.7\%$ of the calculated total width.
Using the QPC model with harmonic oscillator wave functions, we also identified $\rho\rho$ as a dominant mode of the $2\,{}^3F_4$ candidate, together with $a_2(1320)\pi$~\cite{Pang:2015eha}.
In the present calculation, the largest partial width is instead found for $a_2(1700)\pi$.

For the $f'_4(1F)$ assignment at the same input mass, the calculated total width is $375$ MeV.
Its decay pattern is dominated by $K^*\bar K^*$ with a partial width of $102$ MeV.
$K\bar K^*$, $K\bar K_1(1270)$, and $K\bar K_2^*(1430)$ are the next largest modes with partial widths of $68.8$, $43.2$, and $43.0$ MeV, respectively.
These four channels account for approximately $68.5\%$ of the total width.
The $\rho\rho$ and $\omega\omega$ branching fractions are only $0.424\%$ and $0.139\%$, respectively, compared with $11.8\%$ and $3.82\%$ for the $f_4(2F)$ assignment.

 The $f_4(2F)$ width lies within the PDG interval of $250\pm80$ MeV, while the $f'_4(1F)$ width exceeds its upper limit~\cite{ParticleDataGroup:2026mpi}.
The two widths lie on either side of the $310\pm25$ MeV interval obtained in the combined analysis of Ref.~\cite{Anisovich:2000ut}.
The total-width comparison favors the $f_4(2F)$ assignment at the adopted mixing angles.

Both assignments predict $\Gamma_{\rho\rho}/\Gamma_{\omega\omega}\simeq3$, consistent with the WA102 estimate of $2.8\pm0.5$~\cite{WA102:2000bii}.
This ratio therefore does not distinguish their flavor compositions.
The relative strengths of kaon and nonstrange channels provide a clearer test.
For example, at the adopted mixing angles, $\Gamma_{K^*\bar K^*}/\Gamma_{\rho\rho}$ is approximately $0.0251$ for $f_4(2F)$ and $64.3$ for $f'_4(1F)$.
Measurements of these relative decay strengths would directly test the contrasting predictions.
The MGI mass spectrum analysis and the available decay information both favor the nonstrange radial assignment of $f_4(2300)$.
{In fact, most of the final states found in experiments are $n\bar n$, and we also find that when $f_4(2300)$ is taken as $f_4(2F)(n\bar n)$, it decays into $\rho\rho$, $\omega\omega$, $\pi\pi$, and $a_2(1700)\pi$, while $\eta\pi$ is the main decay channel of $a_2(1700)$\cite{Pang:2014laa}, all of these are in good agreement with the experimentally observed decay channels \cite{ParticleDataGroup:2026mpi}.
Therefore, this greatly increases the possibility that $f_4(2300)$ is a candidate state for $f_4(2F)(n\bar n)$.
Of course, the mass of $f_4(1F)(s\bar s)$ is also close to 2320, so it cannot be excluded that the experimental observation of $f_4(2300)$ in the $K\bar K$ final state may be $f_4(1F)(s\bar s)$.}

The strange partner $f'_4(2F)$ is predicted at $2453$ MeV.
As shown in Table~\ref{tab:f4mass}, this mass is $144$ MeV below the GI result and $204$ MeV below the prediction of Ebert \textit{et al.}~\cite{Ebert:2009ub}.
Figure~\ref{fig:pair1F} shows that its total width has a minimum near zero flavor mixing and increases at larger $|\theta_{2\,{}^3F_4}|$.
At the adopted angle $\theta_{2\,{}^3F_4}=6^\circ$ and $\gamma=6.90$, the total width is $216$ MeV.
Varying only $\gamma$ gives a width range of $200$--$233$ MeV.

The dominant decay channels of $f'_4(2F)$ are $K^*\bar K^*(1410)$ and $K^*\bar K^*$, with partial widths of $59.0$ and $39.4$ MeV and branching fractions of $27.3\%$ and $18.2\%$, respectively.
The $K\bar K^*(1410)$ and $K\bar K^*$ modes contribute another $8.91\%$ and $4.83\%$.
The $K^*\bar K_2^*(1430)$, $K\bar K$, and $K\bar K(1460)$ modes each contribute about $4$--$5\%$.
These results suggest searching for $f'_4(2F)$ near $2.45$ GeV in the two leading channels.
Their predicted partial-width ratio is $\Gamma_{K^*\bar K^*(1410)}/\Gamma_{K^*\bar K^*}=1.50$.
The partial-width ratios quoted above are independent of the common pair-creation strength $\gamma$ at fixed masses, wave functions, and mixing angles.

\subsubsection{Predictions for the $3F$, $4F$, $1H$, and $2H$ flavor pairs}

\begin{table*}[tbp]
\caption{Predicted two-body strong-decay widths and branching fractions of $f_4(3F)$, $f'_4(3F)$, $f_4(4F)$, and $f'_4(4F)$.
The calculations use $\gamma=6.90$ and $\theta_{3\,{}^3F_4}=\theta_{4\,{}^3F_4}=6^\circ$.
The adopted parent masses are 2466.29, 2687.91, 2640.48, and 2877.77 MeV, respectively.
Widths are in MeV.
Charge-conjugate modes and all allowed charge combinations are included.
Branching fractions are normalized to the total calculated two-body width of each state.
Channels with $\mathcal B_i\geq1\%$ for at least one state are listed explicitly; the remaining channels are grouped under Other channels.
For daughter mesons labeled by $nL$, $n$ denotes the radial quantum number and $L$ the internal orbital angular momentum.
Dashes denote entries for which no numerical value is quoted, including partial widths below $10^{-3}$ MeV.
Calculated widths and branching fractions are rounded to three significant figures.}
\label{tab:channels_3F_4F_theta3}
\centering
\fontsize{6.5}{7.5}\selectfont\setlength{\tabcolsep}{3pt}\renewcommand{\arraystretch}{0.80}
\begin{ruledtabular}
\begin{tabular}{lrrrrrrrr}
 & \multicolumn{2}{c}{$f_4(3F)$} & \multicolumn{2}{c}{$f'_4(3F)$} & \multicolumn{2}{c}{$f_4(4F)$} & \multicolumn{2}{c}{$f'_4(4F)$} \\
Mode & $\Gamma_i$ & $\mathcal B_i$ (\%) & $\Gamma_i$ & $\mathcal B_i$ (\%) & $\Gamma_i$ & $\mathcal B_i$ (\%) & $\Gamma_i$ & $\mathcal B_i$ (\%) \\
\hline
$a_{2}(3P)\,\pi$ & $28.1$ & $20.4$ & $0.651$ & $0.343$ & $12.5$ & $9.79$ & $0.514$ & $0.283$ \\
$\rho\,\rho$ & $14.1$ & $10.2$ & $0.440$ & $0.232$ & $5.93$ & $4.63$ & $0.304$ & $0.167$ \\
$\pi\,\pi(1300)$ & $11.0$ & $8.01$ & $0.380$ & $0.200$ & $2.47$ & $1.93$ & $0.216$ & $0.119$ \\
$a_{2}(1320)\,\pi$ & $8.63$ & $6.26$ & $0.132$ & $0.0694$ & $6.54$ & $5.10$ & $0.127$ & $0.0697$ \\
$b_{1}(1235)\,\rho$ & $7.59$ & $5.51$ & $0.261$ & $0.138$ & $4.56$ & $3.56$ & $0.214$ & $0.118$ \\
$\pi_{2}(1880)\,\pi$ & $7.00$ & $5.08$ & $0.306$ & $0.161$ & $6.16$ & $4.81$ & $0.191$ & $0.105$ \\
$a_{3}(2030)\,\pi$ & $6.55$ & $4.75$ & $0.236$ & $0.125$ & $2.00$ & $1.56$ & $0.0955$ & $0.0525$ \\
$\pi_{2}(1670)\,\pi$ & $5.65$ & $4.10$ & $0.101$ & $0.0534$ & $3.17$ & $2.47$ & $0.0905$ & $0.0498$ \\
$\pi\,\pi(1800)$ & $4.95$ & $3.59$ & $0.0237$ & $0.0125$ & $4.98$ & $3.89$ & $0.150$ & $0.0824$ \\
$a_{2}(1700)\,\pi$ & $4.64$ & $3.37$ & $0.424$ & $0.224$ & $4.29$ & $3.35$ & $0.214$ & $0.118$ \\
$\rho\,\rho(1450)$ & $3.74$ & $2.71$ & $0.352$ & $0.186$ & $3.98$ & $3.10$ & $0.247$ & $0.136$ \\
$h_{1}(1170)\,\omega$ & $3.59$ & $2.60$ & $0.0946$ & $0.0499$ & $2.23$ & $1.74$ & $0.0795$ & $0.0438$ \\
$a_{1}(1640)\,\pi$ & $3.46$ & $2.51$ & $0.145$ & $0.0764$ & $2.61$ & $2.04$ & $0.142$ & $0.0782$ \\
$\omega\,\omega$ & $3.31$ & $2.40$ & $0.107$ & $0.0566$ & $1.38$ & $1.07$ & $0.0733$ & $0.0403$ \\
$a_{1}(3P)\,\pi$ & $3.13$ & $2.27$ & $0.180$ & $0.0951$ & $0.976$ & $0.762$ & $0.0762$ & $0.0419$ \\
$a_{4}(2F)\,\pi$ & $2.84$ & $2.06$ & $0.206$ & $0.109$ & $2.67$ & $2.09$ & $0.125$ & $0.0686$ \\
$a_{1}(1260)\,\pi$ & $2.43$ & $1.76$ & $0.0560$ & $0.0295$ & $2.03$ & $1.59$ & $0.0298$ & $0.0164$ \\
$a_{4}(2040)\,\pi$ & $2.43$ & $1.76$ & $0.102$ & $0.0537$ & $1.43$ & $1.12$ & $0.0824$ & $0.0454$ \\
$\omega\,\omega(1420)$ & $2.34$ & $1.69$ & $0.123$ & $0.0649$ & $1.98$ & $1.54$ & $0.0825$ & $0.0454$ \\
$a_{2}(4P)\,\pi$ & $2.17$ & $1.57$ & $0.515$ & $0.271$ & $18.7$ & $14.6$ & $0.305$ & $0.168$ \\
$\pi_{4}(1G)\,\pi$ & $1.42$ & $1.03$ & $0.0205$ & $0.0108$ & $0.658$ & $0.514$ & $0.0108$ & $0.00596$ \\
$K^{*}\,\bar K^{*}(1410)$ & $0.0482$ & $0.0350$ & $35.9$ & $18.9$ & $0.119$ & $0.0928$ & $21.5$ & $11.9$ \\
$K^{*}\,\bar K^{*}$ & $0.125$ & $0.0910$ & $23.6$ & $12.5$ & $0.0289$ & $0.0226$ & $15.8$ & $8.72$ \\
$K\,\bar K^{*}(1410)$ & $0.0949$ & $0.0689$ & $20.2$ & $10.6$ & $0.142$ & $0.111$ & $12.6$ & $6.96$ \\
$K^{*}\,\bar K_{2}^{*}(1430)$ & $0.301$ & $0.218$ & $19.7$ & $10.4$ & $0.119$ & $0.0933$ & $18.6$ & $10.2$ \\
$K\,\bar K(1460)$ & $0.0177$ & $0.0129$ & $10.9$ & $5.75$ & $0.0155$ & $0.0121$ & $7.72$ & $4.25$ \\
$K_{1}(1270)\,\bar K^{*}$ & $0.123$ & $0.0895$ & $5.77$ & $3.04$ & $0.0256$ & $0.0200$ & $3.85$ & $2.12$ \\
$K\,\bar K^{*}(3S)$ & \textemdash & \textemdash & $5.91$ & $3.12$ & \textemdash & \textemdash & $4.65$ & $2.56$ \\
$K\,\bar K_{3}^{*}(1780)$ & $0.0126$ & $0.00912$ & $5.31$ & $2.80$ & $0.00158$ & $0.00123$ & $4.35$ & $2.39$ \\
$K\,\bar K_{1}(1400)$ & $0.130$ & $0.0944$ & $4.96$ & $2.62$ & $0.0127$ & $0.00990$ & $2.14$ & $1.18$ \\
$K\,\bar K_{2}(2D)$ & \textemdash & \textemdash & $4.75$ & $2.50$ & \textemdash & \textemdash & $4.71$ & $2.59$ \\
$K\,\bar K(1830)$ & $0.00922$ & $0.00669$ & $4.62$ & $2.44$ & $0.00691$ & $0.00540$ & $6.02$ & $3.31$ \\
$K\,\bar K_{2}(1770)$ & $0.00877$ & $0.00636$ & $4.52$ & $2.39$ & $0.00549$ & $0.00429$ & $2.75$ & $1.51$ \\
$K\,\bar K_{2}^{*}(1430)$ & $0.223$ & $0.162$ & $4.43$ & $2.34$ & $0.0581$ & $0.0454$ & $3.66$ & $2.01$ \\
$K\,\bar K_{1}(1790)$ & \textemdash & \textemdash & $3.52$ & $1.86$ & $0.00938$ & $0.00733$ & $3.75$ & $2.06$ \\
$K\,\bar K$ & $0.00172$ & $0.00125$ & $3.79$ & $2.00$ & \textemdash & \textemdash & $1.11$ & $0.609$ \\
$K^{*}\,\bar K_{1}(1400)$ & $0.00657$ & $0.00477$ & $2.41$ & $1.27$ & $0.00540$ & $0.00422$ & $2.10$ & $1.16$ \\
$K\,\bar K_{3}^{*}(2D)$ & \textemdash & \textemdash & $2.39$ & $1.26$ & \textemdash & \textemdash & $3.57$ & $1.97$ \\
$K\,\bar K_{2}^{*}(2P)$ & \textemdash & \textemdash & $2.25$ & $1.19$ & $0.0170$ & $0.0132$ & $2.20$ & $1.21$ \\
$K\,\bar K'_1(2P)$ & $0.0286$ & $0.0208$ & $2.99$ & $1.58$ & $0.0780$ & $0.0609$ & $4.16$ & $2.29$ \\
$\rho_{3}(1690)\,\rho$ & $0.146$ & $0.106$ & $0.233$ & $0.123$ & $8.95$ & $6.98$ & $0.175$ & $0.0966$ \\
$a_{3}(3F)\,\pi$ & $0.747$ & $0.542$ & $0.507$ & $0.268$ & $6.22$ & $4.86$ & $0.428$ & $0.236$ \\
$\omega_{3}(1670)\,\omega$ & $0.723$ & $0.525$ & $0.0906$ & $0.0478$ & $3.50$ & $2.73$ & $0.0734$ & $0.0404$ \\
$\pi_{2}(3D)\,\pi$ & $0.308$ & $0.224$ & $0.190$ & $0.100$ & $3.20$ & $2.49$ & $0.152$ & $0.0834$ \\
$\pi_{4}(2G)\,\pi$ & \textemdash & \textemdash & $0.00507$ & $0.00267$ & $1.96$ & $1.53$ & $0.0273$ & $0.0150$ \\
$\pi\,\pi(4S)$ & $0.709$ & $0.514$ & $0.260$ & $0.137$ & $1.87$ & $1.46$ & $0.0129$ & $0.00711$ \\
$h_{1}(2P)\,\omega$ & $0.648$ & $0.470$ & $0.0453$ & $0.0239$ & $1.63$ & $1.27$ & $0.0597$ & $0.0329$ \\
$a_{1}(4P)\,\pi$ & \textemdash & \textemdash & $0.0238$ & $0.0126$ & $1.42$ & $1.11$ & $0.130$ & $0.0717$ \\
$K\,\bar K_{2}(2250)$ & \textemdash & \textemdash & \textemdash & \textemdash & \textemdash & \textemdash & $8.02$ & $4.41$ \\
$K^{*}\,\bar K_{3}^{*}(1780)$ & \textemdash & \textemdash & $0.932$ & $0.492$ & \textemdash & \textemdash & $7.62$ & $4.19$ \\
$K\,\bar K_{4}^{*}(1F)$ & \textemdash & \textemdash & $1.47$ & $0.777$ & $0.00230$ & $0.00180$ & $1.89$ & $1.04$ \\
$K\,\bar K_{1}(1270)$ & $0.00998$ & $0.00724$ & $1.62$ & $0.855$ & $0.0332$ & $0.0259$ & $1.86$ & $1.02$ \\
$K\,\bar K'_1(3P)$ & \textemdash & \textemdash & $0.00207$ & $0.00109$ & \textemdash & \textemdash & $1.85$ & $1.02$ \\
\hline
Other channels & $4.33$ & $3.14$ & $11.4$ & $6.03$ & $7.37$ & $5.75$ & $30.7$ & $16.9$ \\
Total & $138$ & $100$ & $190$ & $100$ & $128$ & $100$ & $182$ & $100$ \\
\end{tabular}
\end{ruledtabular}
\end{table*}

\begin{table*}[tbp]
\caption{Predicted two-body strong-decay widths and branching fractions of $f_4(1H)$, $f'_4(1H)$, $f_4(2H)$, and $f'_4(2H)$.
The calculations use $\gamma=6.90$ and $\theta_{1\,{}^3H_4}=\theta_{2\,{}^3H_4}=0^\circ$.
The adopted parent masses are 2405.54, 2606.79, 2588.92, and 2801.72 MeV, respectively.
Widths are in MeV.
Charge-conjugate modes and all allowed charge combinations are included.
Branching fractions are normalized to the total calculated two-body width of each state.
Channels with $\mathcal B_i\geq1\%$ for at least one state are listed explicitly; the remaining channels are grouped under Other channels.
The daughter-state notation follows Table~\ref{tab:channels_3F_4F_theta3}.
Dashes denote entries for which no numerical value is quoted, including partial widths below $10^{-3}$ MeV.
Calculated widths and branching fractions are rounded to three significant figures.}
\label{tab:channels_1H_2H_theta0}
\centering
\scriptsize\setlength{\tabcolsep}{3pt}\renewcommand{\arraystretch}{0.80}
\begin{ruledtabular}
\begin{tabular}{lrrrrrrrr}
 & \multicolumn{2}{c}{$f_4(1H)$} & \multicolumn{2}{c}{$f'_4(1H)$} & \multicolumn{2}{c}{$f_4(2H)$} & \multicolumn{2}{c}{$f'_4(2H)$} \\
Mode & $\Gamma_i$ & $\mathcal B_i$ (\%) & $\Gamma_i$ & $\mathcal B_i$ (\%) & $\Gamma_i$ & $\mathcal B_i$ (\%) & $\Gamma_i$ & $\mathcal B_i$ (\%) \\
\hline
$a_{3}(1875)\,\pi$ & $110$ & $29.5$ & \textemdash & \textemdash & $39.3$ & $14.3$ & \textemdash & \textemdash \\
$\pi_{2}(1670)\,\pi$ & $62.8$ & $16.8$ & \textemdash & \textemdash & $33.4$ & $12.1$ & \textemdash & \textemdash \\
$\pi_{4}(1G)\,\pi$ & $58.6$ & $15.7$ & \textemdash & \textemdash & $9.58$ & $3.48$ & \textemdash & \textemdash \\
$\rho\,\rho$ & $28.1$ & $7.51$ & \textemdash & \textemdash & $2.62$ & $0.951$ & \textemdash & \textemdash \\
$b_{1}(1235)\,\rho$ & $18.9$ & $5.04$ & \textemdash & \textemdash & $7.30$ & $2.65$ & \textemdash & \textemdash \\
$\pi_{2}(1880)\,\pi$ & $15.8$ & $4.22$ & \textemdash & \textemdash & $7.27$ & $2.64$ & \textemdash & \textemdash \\
$\eta(1S)\,\eta_{2}(1D)$ & $10.2$ & $2.73$ & \textemdash & \textemdash & $4.24$ & $1.54$ & \textemdash & \textemdash \\
$a_{1}(1260)\,\pi$ & $8.34$ & $2.23$ & \textemdash & \textemdash & $7.18$ & $2.61$ & \textemdash & \textemdash \\
$a_{2}(1320)\,\pi$ & $7.72$ & $2.07$ & \textemdash & \textemdash & $4.66$ & $1.69$ & \textemdash & \textemdash \\
$h_{1}(1170)\,\omega$ & $7.24$ & $1.94$ & \textemdash & \textemdash & $3.11$ & $1.13$ & \textemdash & \textemdash \\
$a_{4}(2040)\,\pi$ & $6.65$ & $1.78$ & \textemdash & \textemdash & $2.29$ & $0.831$ & \textemdash & \textemdash \\
$\omega\,\omega$ & $6.63$ & $1.77$ & \textemdash & \textemdash & $0.602$ & $0.219$ & \textemdash & \textemdash \\
$a_{1}(1640)\,\pi$ & $6.35$ & $1.70$ & \textemdash & \textemdash & $0.458$ & $0.166$ & \textemdash & \textemdash \\
$a_{3}(2030)\,\pi$ & $6.28$ & $1.68$ & \textemdash & \textemdash & $37.5$ & $13.6$ & \textemdash & \textemdash \\
$h_{1}(1170)\,h_{1}(1170)$ & $3.85$ & $1.03$ & $0.00165$ & $3.01\times10^{-4}$ & $2.18$ & $0.793$ & \textemdash & \textemdash \\
$K\,\bar K'_2(1D)$ & \textemdash & \textemdash & $118$ & $21.5$ & \textemdash & \textemdash & $59.8$ & $15.3$ \\
$K\,\bar K'_3(1F)$ & \textemdash & \textemdash & $95.2$ & $17.3$ & \textemdash & \textemdash & $50.5$ & $12.9$ \\
$K_{1}(1270)\,\bar K^{*}$ & $0.0388$ & $0.0104$ & $70.3$ & $12.8$ & $0.00469$ & $0.00171$ & $35.6$ & $9.09$ \\
$K^{*}\,\bar K_{1}(1400)$ & \textemdash & \textemdash & $45.1$ & $8.22$ & $0.00212$ & $7.71\times10^{-4}$ & $21.0$ & $5.36$ \\
$K^{*}\,\bar K^{*}$ & $0.0370$ & $0.00989$ & $34.5$ & $6.28$ & \textemdash & \textemdash & $15.9$ & $4.06$ \\
$K\,\bar K_{1}(1400)$ & \textemdash & \textemdash & $31.7$ & $5.77$ & \textemdash & \textemdash & $27.0$ & $6.90$ \\
$K^{*}\,\bar K_{2}^{*}(1430)$ & $0.00393$ & $0.00105$ & $16.5$ & $3.01$ & $0.00388$ & $0.00141$ & $12.4$ & $3.17$ \\
$K^{*}\,\bar K^{*}(1410)$ & \textemdash & \textemdash & $16.3$ & $2.96$ & $0.0163$ & $0.00591$ & $17.2$ & $4.39$ \\
$K\,\bar K_{1}(1790)$ & $0.0257$ & $0.00686$ & $11.5$ & $2.09$ & $0.201$ & $0.0729$ & $1.43$ & $0.365$ \\
$K\,\bar K_{3}^{*}(1780)$ & $0.147$ & $0.0394$ & $15.2$ & $2.77$ & $0.0117$ & $0.00425$ & $6.26$ & $1.60$ \\
$K\,\bar K_{2}^{*}(1430)$ & $0.209$ & $0.0559$ & $14.2$ & $2.59$ & $0.0267$ & $0.00969$ & $8.31$ & $2.12$ \\
$\eta(1S)\,\eta'_2(1D)$ & \textemdash & \textemdash & $11.2$ & $2.05$ & \textemdash & \textemdash & $3.25$ & $0.830$ \\
$K\,\bar K'_1(2P)$ & $0.00105$ & $2.81\times10^{-4}$ & $14.2$ & $2.59$ & $0.0206$ & $0.00750$ & $0.111$ & $0.0283$ \\
$K\,\bar K(1460)$ & $0.0289$ & $0.00773$ & $7.82$ & $1.42$ & $0.00846$ & $0.00307$ & $2.68$ & $0.686$ \\
$\pi_{4}(2G)\,\pi$ & \textemdash & \textemdash & \textemdash & \textemdash & $55.4$ & $20.1$ & \textemdash & \textemdash \\
$\pi_{2}(3D)\,\pi$ & \textemdash & \textemdash & \textemdash & \textemdash & $8.21$ & $2.98$ & \textemdash & \textemdash \\
$\rho\,\rho(1450)$ & $0.369$ & $0.0988$ & \textemdash & \textemdash & $6.45$ & $2.35$ & \textemdash & \textemdash \\
$a_{3}(3F)\,\pi$ & \textemdash & \textemdash & \textemdash & \textemdash & $6.29$ & $2.29$ & \textemdash & \textemdash \\
$a_{4}(2F)\,\pi$ & $0.0281$ & $0.00751$ & \textemdash & \textemdash & $3.87$ & $1.41$ & \textemdash & \textemdash \\
$a_{1}(3P)\,\pi$ & $0.0803$ & $0.0215$ & \textemdash & \textemdash & $3.76$ & $1.37$ & \textemdash & \textemdash \\
$a_{5}(1H)\,\pi$ & \textemdash & \textemdash & \textemdash & \textemdash & $3.44$ & $1.25$ & \textemdash & \textemdash \\
$a_{2}(3P)\,\pi$ & $0.102$ & $0.0274$ & \textemdash & \textemdash & $3.42$ & $1.24$ & \textemdash & \textemdash \\
$\omega\,\omega(1420)$ & $0.266$ & $0.0712$ & \textemdash & \textemdash & $3.04$ & $1.11$ & \textemdash & \textemdash \\
$K\,\bar K'_2(2D)$ & \textemdash & \textemdash & \textemdash & \textemdash & \textemdash & \textemdash & $37.7$ & $9.62$ \\
$K^{*}(1410)\,\bar K_{1}(1270)$ & \textemdash & \textemdash & \textemdash & \textemdash & \textemdash & \textemdash & $9.39$ & $2.40$ \\
$K^{*}\,\bar K_{1}(1790)$ & \textemdash & \textemdash & \textemdash & \textemdash & \textemdash & \textemdash & $8.65$ & $2.21$ \\
$K_{1}(1270)\,\bar K_{1}(1270)$ & \textemdash & \textemdash & $4.73$ & $0.862$ & \textemdash & \textemdash & $6.40$ & $1.64$ \\
$K\,\bar K_{3}^{*}(2D)$ & \textemdash & \textemdash & \textemdash & \textemdash & \textemdash & \textemdash & $5.63$ & $1.44$ \\
$K\,\bar K_{1}(3P)$ & \textemdash & \textemdash & \textemdash & \textemdash & \textemdash & \textemdash & $4.12$ & $1.05$ \\
$K\,\bar K'_1(3P)$ & \textemdash & \textemdash & \textemdash & \textemdash & \textemdash & \textemdash & $6.39$ & $1.63$ \\
\hline
Other channels & $14.7$ & $3.94$ & $42.5$ & $7.73$ & $19.3$ & $7.03$ & $51.7$ & $13.2$ \\
Total & $374$ & $100$ & $549$ & $100$ & $275$ & $100$ & $391$ & $100$ \\
\end{tabular}
\end{ruledtabular}
\end{table*}

\begin{figure*}[tbp]
\includegraphics[width=\textwidth]{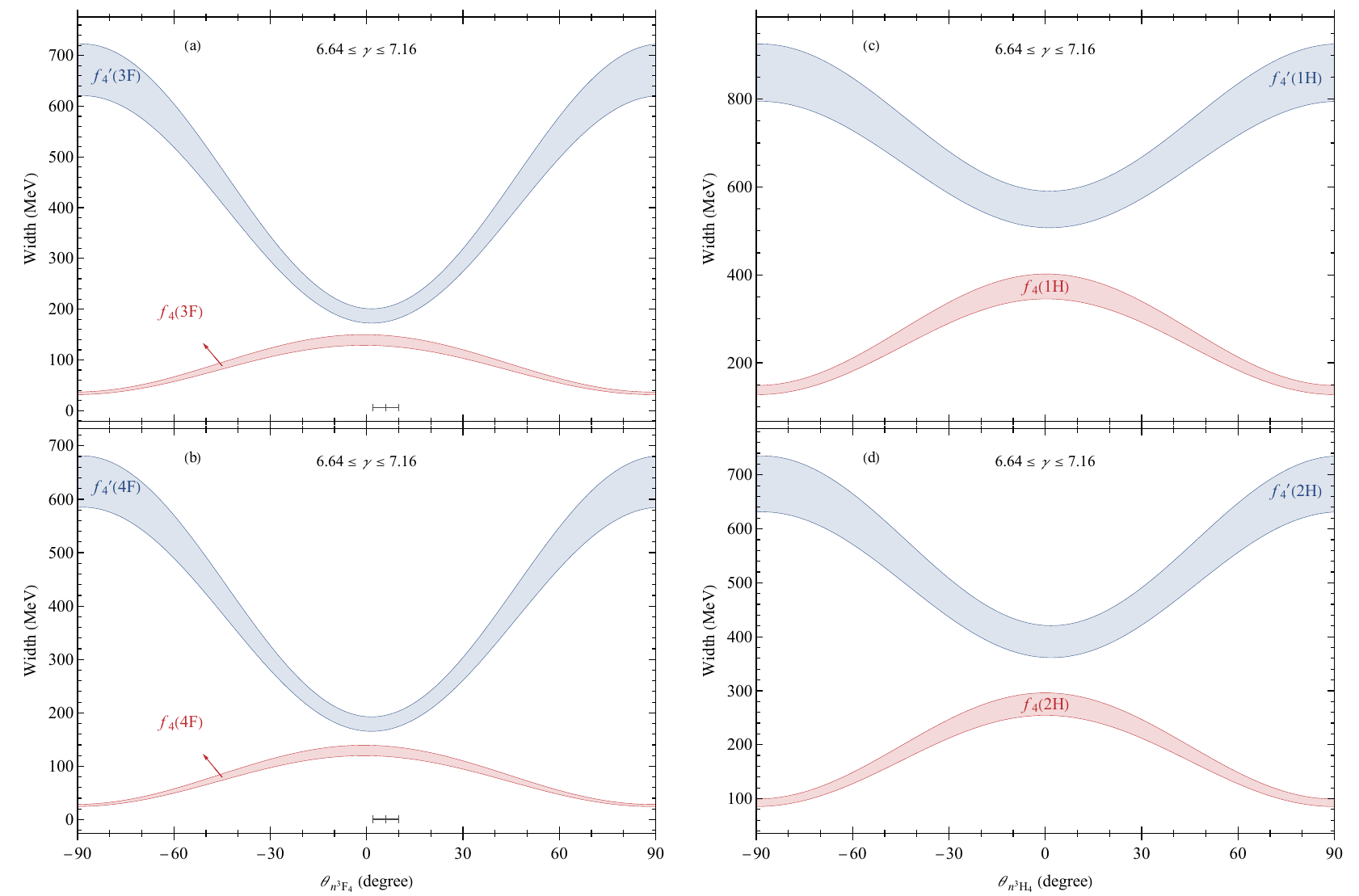}
\caption{Calculated total decay widths of the flavor pairs as functions of their respective parent flavor-mixing angles: (a) $3F$, (b) $4F$, (c) $1H$, and (d) $2H$.
Red and blue bands represent the unprimed and primed states, respectively.
At each angle, the bands are obtained by varying $\gamma$ from $6.64$ to $7.16$, with all other inputs fixed.
The parent masses are those used in Tables~\ref{tab:channels_3F_4F_theta3} and~\ref{tab:channels_1H_2H_theta0}.
The brackets in panels (a) and (b) indicate the adopted $F$-wave interval from $2^\circ$ to $10^\circ$, with the middle tick at $6^\circ$.}
\label{fig:34F12H}
\end{figure*}

{ We now extend the QPC analysis to the $3F$, $4F$, $1H$, and $2H$ flavor pairs.
The calculated OZI-allowed two-body partial widths and branching fractions are presented in Tables~\ref{tab:channels_3F_4F_theta3},~\ref{tab:channels_1H_2H_theta0} and Figure~\ref{fig:34F12H}.

The nonstrange state $f_4(3F)$ is predicted at $2466$ MeV, with a total two-body strong-decay width of $138$ MeV.
Its leading channels are $a_2(3P)\pi$, $\rho\rho$, $\pi\pi(1300)$, and $a_2(1320)\pi$, with branching fractions of $20.4\%$, $10.2\%$, $8.01\%$, and $6.26\%$, respectively.

Its strange partner $f'_4(3F)$ is predicted at $2688$ MeV, with a total width of $190$ MeV.
$K^*\bar K^*(1410)$, $K^*\bar K^*$, $K\bar K^*(1410)$, and $K^*\bar K_2^*(1430)$ are its important decay channels.
Their branching fractions are $18.9\%$, $12.5\%$, $10.6\%$, and $10.4\%$, respectively.
Together, these modes account for $52.4\%$ of the total width.
This preference for kaon final states contrasts with the leading nonstrange channels of $f_4(3F)$.

For $f_4(4F)$, the predicted mass and total width are $2640$ and $128$ MeV, respectively.
The largest contributions come from $a_2(4P)\pi$ and $a_2(3P)\pi$, with branching fractions of $14.6\%$ and $9.79\%$.
The $\rho_3(1690)\rho$ channel also contributes $6.98\%$.
Compared with $f_4(3F)$, the $a_2(4P)\pi$ branching fraction increases from $1.57\%$ to $14.6\%$, providing a useful distinction between the predicted decay patterns of the two radial excitations.

Its strange partner $f'_4(4F)$ is predicted at $2878$ MeV, with a total width of $182$ MeV.
The leading channels remain $K^*\bar K^*(1410)$, $K^*\bar K_2^*(1430)$, $K^*\bar K^*$, and $K\bar K^*(1410)$, with branching fractions of $11.9\%$, $10.2\%$, $8.72\%$, and $6.96\%$, respectively.

The nonstrange $f_4(1H)$ is predicted at $2406$ MeV, with a total width of $374$ MeV.
The most important decay mode of $f_4(1H)$ is $a_3(1875)\pi$ with a branching fraction of $29.5\%$.
$\pi_2(1670)\pi$ and $\pi_4(1G)\pi$ also make sizable contributions to the total width, with branching fractions of $16.8\%$ and $15.7\%$, respectively.
Together, these channels account for approximately $62\%$ of the calculated total width.
The $\rho\rho$, $b_1(1235)\rho$, and $\pi_2(1880)\pi$ modes provide further contributions of $7.51\%$, $5.04\%$, and $4.22\%$, respectively.

The strange partner $f'_4(1H)$ is predicted at $2607$ MeV, with a total width of $549$ MeV.
The first three decay channels are $K\bar K'_2(1D)$, $K\bar K'_3(1F)$ and $K_1(1270)\bar K^*$, with branching fractions of $21.5\%$, $17.3\%$, and $12.8\%$, which contribute about half of the total width.

The first radial excitation of the nonstrange $H$-wave state, $f_4(2H)$, is predicted at $2589$ MeV.
Its total width is $275$ MeV, smaller than that of $f_4(1H)$ despite the higher mass.
The most important decay channel is $\pi_4(2G)\pi$, with a width and branching fraction of 55.4 MeV and $20.1\%$, respectively.
$a_3(1875)\pi$, $a_3(2030)\pi$, and $\pi_2(1670)\pi$ are its important decay modes, with branching fractions of $14.3\%$, $13.6\%$, and $12.1\%$, respectively.

The strange $f'_4(2H)$ is predicted at $2802$ MeV, with a total width of $391$ MeV.
The $K\bar K'_2(1D)$ and $K\bar K'_3(1F)$ channels remain prominent, with branching fractions of $15.3\%$ and $12.9\%$, respectively.
The $K\bar K'_2(2D)$ and $K_1(1270)\bar K^*$ modes contribute $9.62\%$ and $9.09\%$.

Figure~\ref{fig:34F12H} shows the dependence of the total widths on the parent flavor-mixing angles.
In all four panels, the unprimed-state widths have maxima near zero mixing, whereas the primed-state widths have minima in this region.
The primed widths increase as their nonstrange components grow towards large mixing angles.
At each angle, the bands are obtained by varying $\gamma$ from $6.64$ to $7.16$, with all other inputs fixed.
At the reference angles, this variation gives total-width intervals of $507$--$591$ MeV and $362$--$421$ MeV for $f'_4(1H)$ and $f'_4(2H)$, respectively.
Both states remain broad throughout this range.

The flavor composition can also be tested through nonstrange decay channels.
At zero parent flavor mixing, the partial widths of the strange $H$-wave states into the purely nonstrange $\rho\rho$ and $a_J\pi$ channels vanish in the present OZI-allowed calculation.
Measurements of substantial branching fractions into these modes would therefore test the assumed pure $s\bar s$ composition.

The predicted $F$- and $H$-wave states lie close in mass but differ substantially in their widths and decay patterns.
The nonstrange $1H$ and $2H$ states lie about $60$ and $51$ MeV below the $3F$ and $4F$ states, respectively; the corresponding separations for the strange partners are $81$ and $76$ MeV.
At the adopted reference angles, each of these $H$-wave states has more than twice the width of its nearby $F$-wave counterpart.
The large $a_3\pi$ and $\pi_4\pi$ contributions in the nonstrange $H$ sector, and the prominent $K\bar K'_2$, $K\bar K'_3$, and $K_1\bar K^*$ modes in the strange $H$ sector, provide useful channels for distinguishing these orbital assignments.
A combined analysis of the leading decay channels and their angular distributions would help separate the nearby $F$- and $H$-wave states.
}

\section{Conclusion}\label{sec4}

We have investigated the mass spectrum and OZI-allowed two-body strong decays of the $f_4$ family within the conventional $q\bar q$ framework.
The MGI model provides the masses and numerical spatial wave functions used in the QPC decay calculations.
The predicted $F$-wave radial level spacings decrease with increasing excitation, while several $F$- and $H$-wave levels lie close in mass.
Their decay properties therefore provide complementary information for identifying the members of the family.

Taking $f_4(2050)$ as the mainly nonstrange $1\,{}^3F_4$ reference state, we determine the common pair-creation strength from its measured total width.
Using the experimental parent mass and the reference mixing angle $\theta_{1\,{}^3F_4}=6^\circ$ gives $\gamma=6.90^{+0.26}_{-0.27}$.
The predicted leading channels are $\rho\rho$, $a_2(1320)\pi$, $\pi\pi$, and $\omega\omega$, with $\rho\rho$ accounting for $44.2\%$ of the calculated total width.
Our theoretical results are basically consistent with the experimental data.

Among the two $F$-wave assignments examined for $f_4(2300)$, the MGI mass spectrum favors the $2\,{}^3F_4$ configuration with a dominant $n\bar n$ component over the $1\,{}^3F_4$ configuration with a dominant $s\bar s$ component.
At the common input mass of $2320$ MeV, their calculated total widths are $232$ and $375$ MeV, respectively.
The PDG width favors the $2\,{}^3F_4$ assignment at the adopted mixing angles.
The two configurations predict markedly different decay patterns.
The decays of the strange partner of $f_4(2050)$ are dominated by $K^*\bar K^*$ and $K\bar K^*$ which can help experiments search for this state.

The strange $2F$ partner is predicted at $2453$ MeV, with a total width of $216$ MeV at $\theta_{2\,{}^3F_4}=6^\circ$ and $\gamma=6.90$.
Its leading channels are $K^*\bar K^*(1410)$ and $K^*\bar K^*$, with a predicted partial-width ratio of $1.50$, suggesting searches in these channels near $2.45$ GeV.
The higher $3F$ and $4F$ flavor pairs lie between $2466$ and $2878$ MeV.
At the same reference mixing angle and pair-creation strength, their total two-body widths range from $128$ to $190$ MeV.
The nonstrange states have prominent $a_2(3P)\pi$ and $a_2(4P)\pi$ modes, while their strange partners favor channels involving kaon resonances, including $K^*\bar K^*(1410)$, $K^*\bar K^*$, and $K^*\bar K_2^*(1430)$.

The $1H$ and $2H$ flavor pairs are predicted between $2406$ and $2802$ MeV.
At zero flavor mixing and $\gamma=6.90$, their total two-body widths range from $275$ to $549$ MeV.
Although their masses lie close to those of the $3F$ and $4F$ states, each calculated $H$-wave width exceeds twice that of its nearby $F$-wave counterpart in the same flavor sector.
The nonstrange $1H$ and $2H$ states have their largest contributions in $a_3(1875)\pi$ and $\pi_4(2G)\pi$, respectively.
For the strange $H$-wave states, $K\bar K'_2(1D)$, $K\bar K'_3(1F)$, and $K_1(1270)\bar K^*$ are important search channels.
These differences in widths and leading decay modes provide useful criteria for distinguishing nearby orbital excitations.

At fixed masses, spatial wave functions, and mixing angles, the predicted partial-width ratios are independent of the common pair-creation strength.
Measurements of these ratios and angular distributions in the leading channels would help distinguish the higher radial and orbital excitations of the $f_4$ family.

\section{ACKNOWLEDGMENTS}
This work is supported by the National Natural Science Foundation of China under Grant No.~11965016 and No.~12047501, and by the Natural Science Foundation of Qinghai Province under Grant No.~2022-ZJ-939Q. Zhe-Qi Huang, Lei Huang, and Yu-Qiang Li contributed equally to this work.

\bibliographystyle{apsrev4-2}
\bibliography{hepref_INSPIRE}

\end{document}